\ifdefined\draft
\documentclass{article}
\usepackage[top=1in,bottom=1in,left=1in,right=1in]{geometry}
\else
\documentclass[journal]{IEEEtran}
\fi
\usepackage{algorithm}
\usepackage{algpseudocode}
\usepackage{url}
\usepackage{cite}
\usepackage{graphicx}

\usepackage[cmex10]{amsmath}
\usepackage{amsfonts}
\usepackage{array}

\usepackage{mdwmath}
\usepackage{mdwtab}
\usepackage{eqparbox}

\usepackage[tight,footnotesize]{subfigure}
\usepackage[font=footnotesize]{subfig}

\usepackage{color}

\usepackage{stfloats}
\title{Learning the Treatment Effects on FTIR Signals Subject to Multiple Sources of Uncertainties}

\ifdefined\draft

\author{}

\else
\author{Hongzhen Tian,~
        Andi Wang,~
        Jialei Chen,~
        Xuzhou Jiang,~
        Jianjun Shi,~\\
        Chuck Zhang,~
        Yajun Mei,~
        and~Ben Wang,~
\thanks{Corresponding author: Chuck Zhang.}
\thanks{H. Tian, A. Wang, and J. Chen are with the H. Milton Stewart School of Industrial and Systems Engineering, Georgia Institute of Technology, Atlanta, GA 30332 USA (e-mail: hongzhentian@gatech.edu; andi.wang@gatech.edu; jialei.chen@gatech.edu).}
\thanks{X. Jiang is with the Georgia Tech Manufacturing Institute, Georgia Institute of Technology, Atlanta, GA 30332 USA (e-mail: xjiang@gatech.edu)}
\thanks{J. Shi is with the H. Milton Stewart School of Industrial and Systems Engineering and the George W. Woodruff School of Mechanical Engineering, Georgia Institute of Technology, Atlanta, GA 30332 USA (e-mail: jianjun.shi@isye.gatech.edu).}
\thanks{C. Zhang is with the H. Milton Stewart School of Industrial and Systems Engineering and the Georgia Tech Manufacturing Institute, Georgia Institute of Technology, Atlanta, GA 30332 USA (e-mail: chuck.zhang@gatech.edu).}
\thanks{Y. Mei is with the H. Milton Stewart School of Industrial and Systems Engineering, Georgia Institute of Technology, Atlanta, GA 30332 USA (e-mail: yajun.mei@isye.gatech.edu).}
\thanks{B. Wang is with the H. Milton Stewart School of Industrial and Systems Engineering, the Georgia Tech Manufacturing Institute, and the School of Materials Science and Engineering, Georgia Institute of Technology, Atlanta, GA 30332 USA (e-mail: ben.wang@gatech.edu). }
}
\fi
\begin{document}

\markboth{IEEE TRANSACTIONS ON AUTOMATION SCIENCE AND ENGINEERING}
{Tian \MakeLowercase{\textit{et al.}}: Learning the Treatment Effects on FTIR Signals Subject to Multiple Sources of Uncertainties}


\maketitle

\begin{abstract}
   Fourier-transform infrared spectroscopy (FTIR) is a versatile technique for characterizing the chemical composition of the various uncertainties, including baseline shift and multiplicative error.
   This study aims at analyzing the effect of certain treatment on the FTIR responses subject to these uncertainties. A two-step method is proposed to quantify the treatment effect on the FTIR signals. First, an optimization problem is solved to calculate the template signal by aligning the pre-treatment FTIR signals. Second, the effect of treatment is decomposed  as  the pattern of modification $\mathbf{g}$ that describes the overall treatment effect on the spectra and a vector of effect $\boldsymbol{\delta}$ that describes the degree of modification. $\mathbf g$ and $\boldsymbol{\delta}$ are solved by another optimization problem. They have explicit engineering interpretations and provide useful information on how the treatment effect change the surface chemical components.
   The effectiveness of the proposed method is first validated in a simulation. In a real case study, it's used to investigate how the plasma exposure applied at various heights affects the FTIR signal which indicates the change of the chemical composition on the composite material. The vector of effects indicates the range of effective plasma height, and the pattern of modification matches existing engineering knowledge well.  

\end{abstract}

\def\abstractname{Note to Practitioners}
\begin{abstract}

Fourier-transform infrared spectroscopy is frequently used to characterize the surface chemical composition of a material. Due to the large uncertainty of the FTIR signals,  they are usually observed visually by experienced engineers and technicians in industrial applications. Drawing conclusions by observing FTIR signals can be both time-consuming and inaccurate.
This study is motivated by the problem of understanding the effect of plasma exposure on the surface property of carbon fiber reinforced polymer material. In this study, we investigated the uncertainties associated with FTIR signals, and proposed a general, systematic method to quantify the effect of surface treatment on the FTIR signals that are subject to offset and scaling error. A two-step analytic procedure is proposed, which provides the information on how the plasma exposure distorts the FTIR signal, and how the plasma height relates to the magnitude of the distortion. 
The methodology in this article can be used to analyze the treatment effect on a variety of spectroscopic measurements that are subject to uncertainties like offset shift and multiplicative errors. As the offset and scaling error significantly affect the measurements collected from portable spectrometers in the manufacturing environment,  this study can expand the applications of in-situ hand-held spectrometer metrology in manufacture industries.



\end{abstract}

\ifdefined\draft
\else
\fi
%
\section{Introduction}

%
%
%
%

\ifdefined\draft
Fourier-transform 
\else
\IEEEPARstart{F}{ourier-transform} 
\fi
infrared spectroscopy (FTIR) measurements provide a sensitive, non-destructive way of understanding the material's chemical composition. Like most structural spectroscopic techniques, it characterizes the chemical composition of the material by the absorbance/reflectance of light in a range of frequencies \cite{rein2014analysis,bassan2011light,ctucureanu2016ftir,duvall2010detection}. Specifically, FTIR spectrometer shines a beam of infrared light that contains a range of spectral components on the sample, and measures the intensity of the absorbed light at every frequency. As each chemical bond in the material only absorbs the infrared radiation at its characteristic frequency, the absorbance intensity of each light component indicates the richness of corresponding chemical bond in the examined sample \cite{ctucureanu2016ftir,afseth2012extended,bassan2011light}. The measurement signal obtained from FTIR spectrometer is called the FTIR spectrum. It takes the form of a high dimensional vector, denoting the intensity of infrared absorption at a sequence of frequencies.

The Carbon Fiber Reinforced Polymer (CFRP) composite grabs increasing attention in the aerospace industry for its lightweight, excellent strength and other properties. However, safety concerns on its bonding quality are essential due to low surface free energy \cite{han2014evaluation}. 
To improve bonding quality, surface modification methods have been well-developed to improve CFRP surface energy, including thermal treatment, wet chemical or electrochemical oxidation, plasma treatment, gas-phase oxidation, coating treatment, and so on \cite{sharma2014carbon,tiwari2014surface,kakhki2014review}. 

Among these surface modification methods, plasma exposure is one of the most popular ones with its special advantages
\cite{waters2017effect,de2008surface,kwon2005surface,mukhopadhyay2002plasma,liston1993plasma}. It is a non-destructive method allowing greater control over the number of unwanted reaction pathways \cite{sharma2014carbon
}.  
Previous research indicated that plasma exposure can increase the wettability of the material by modifying the chemical composition  and the physical structures of its surface layer \cite{de2008surface,kwon2005surface,liston1993plasma,tiwari2014surface}, which in turn improves the bonding quality between CFRP panels in aerospace applications.
However, it remains unclear how the plasma height, the distance between the plasma nozzle and the sample, affects the chemical components of CFRP indicated by FTIR measurements.

\begin{figure}
    \centering
    \includegraphics[width=3.5 in]{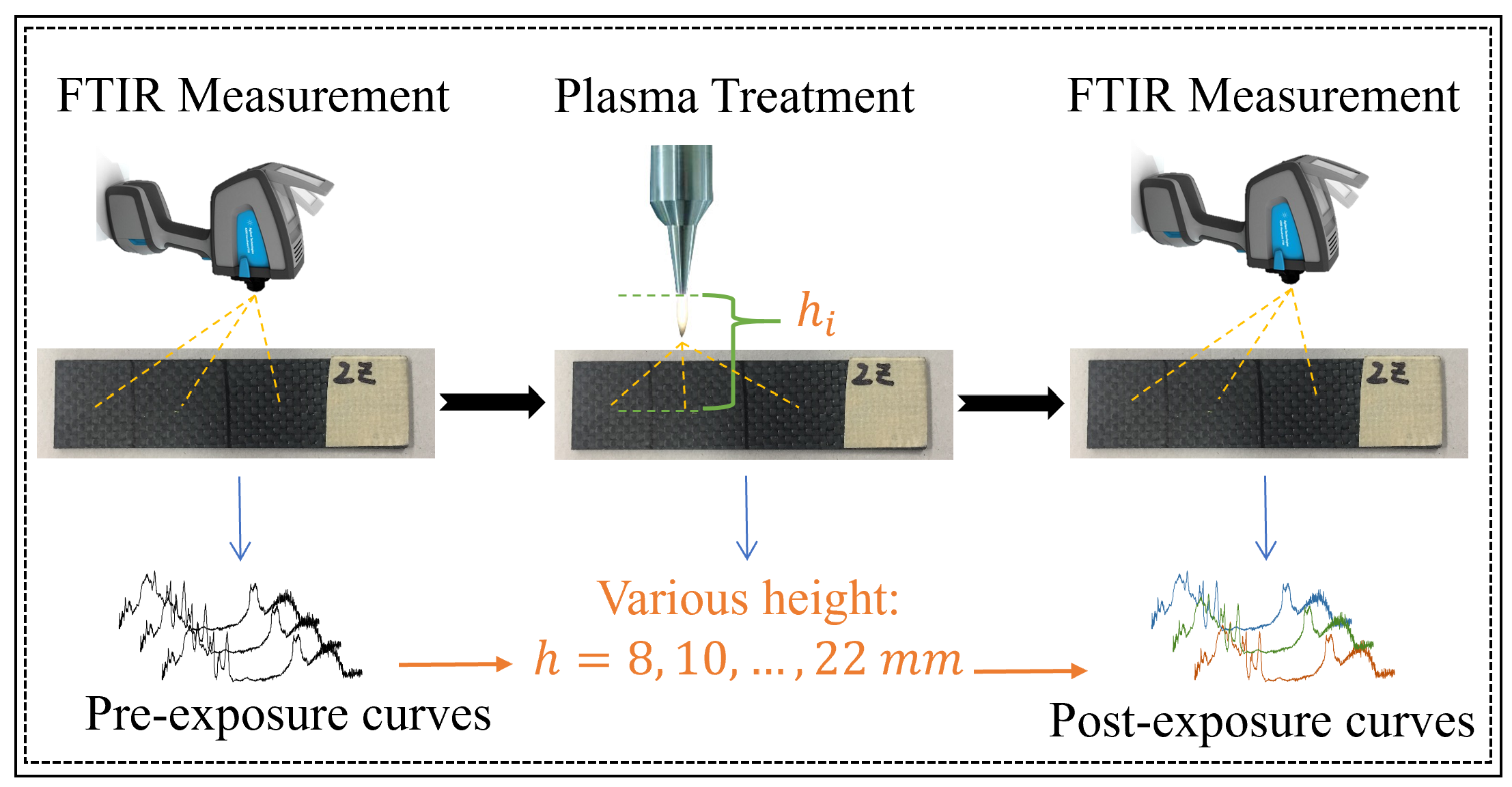}
    \newline
    
    \caption{The experimental setup. The CFRP coupons were processed by plasma, and the plasma height is set at $h_1,...,h_n.$ The FTIR measurements were taken before and after the plasma exposure to capture the change of chemical components on the CFRP surface.}
    \label{fig1}
\end{figure}

Motivated by the problem of quantifying the effect of plasma exposure on the CFRP material when plasma height varies, this study aims at tackling a more general problem of understanding the surface treatment effect of various strength on FTIR signals. Specifically, consider the experiment detailed in Fig.~\ref{fig1}, where we collect the pre-exposure FTIR measurements and post-exposure FTIR signal on a number of CFRP coupons, where the strength of treatment, described by the plasma height, is set to prescribed values from 8mm to 22mm. From the data, we seek to understand how the plasma height affects the FTIR measurements taken on a sample surface before and after the plasma exposure at different heights, which will shed some light on how the chemical composition of CFRP is changed by the plasma treatment. 

Modeling the treatment effect from FTIR signals involves two major challenges. First, FTIR spectra are subject to many uncertainty sources, including light scattering, optical path length variations, and temperature variation \cite{afseth2012extended,bassan2011light,cornel2008quantitative}. 
 Besides, the measurement uncertainties of FTIR spectra have been widely recognized. In industrial applications, the FTIR signals are usually observed visually by experienced engineers and technicians \cite{rein2014analysis,vsedvenkova2008thermal,THowie2017Thermal} and such approaches can be both time-consuming and sometimes inaccurate. 
 These uncertainties result in the offset shift and the multiplicative error, and the latter also influences the magnitude of the noise on the FTIR measurements. 
 To decrease the level of uncertainty in FTIR measurements, the preprocessing of the FTIR signals is studied in a series of literature. For example, Cornel et al. \cite{cornel2008quantitative} reviewed multiple preprocessing procedures for analyzing the FTIR signals. However, most of them are ad-hoc methods that do not fully characterize the sources of uncertainty and involved error patterns in the FTIR spectra. Among them, the only exception is the multiplicative scatter correction (MSC) method, which characterized the multiplicative error and offset shift of the FTIR spectra. However, the MSC model fails to consider the magnitude-dependent noise based on the FTIR spectra.

The second challenge is that surface treatment usually  has a nonlinear effect on the FTIR signal. For example, the effect of the plasma exposure is nonlinear in general \cite{mukhopadhyay2002plasma}, as when the plasma height is small the effect of adjusting plasma height for one unit is very significant, whereas when the plasma height is large, changing the plasma height for one unit will barely affect the FTIR signal. Due to the nonlinearity, the univariate effect on the high-dimensional measurements cannot be characterized through existing methods  such as the functional linear models \cite{ramsay2005applied}.

\begin{figure}
    \centering
  
    \includegraphics[width=3.5 in]{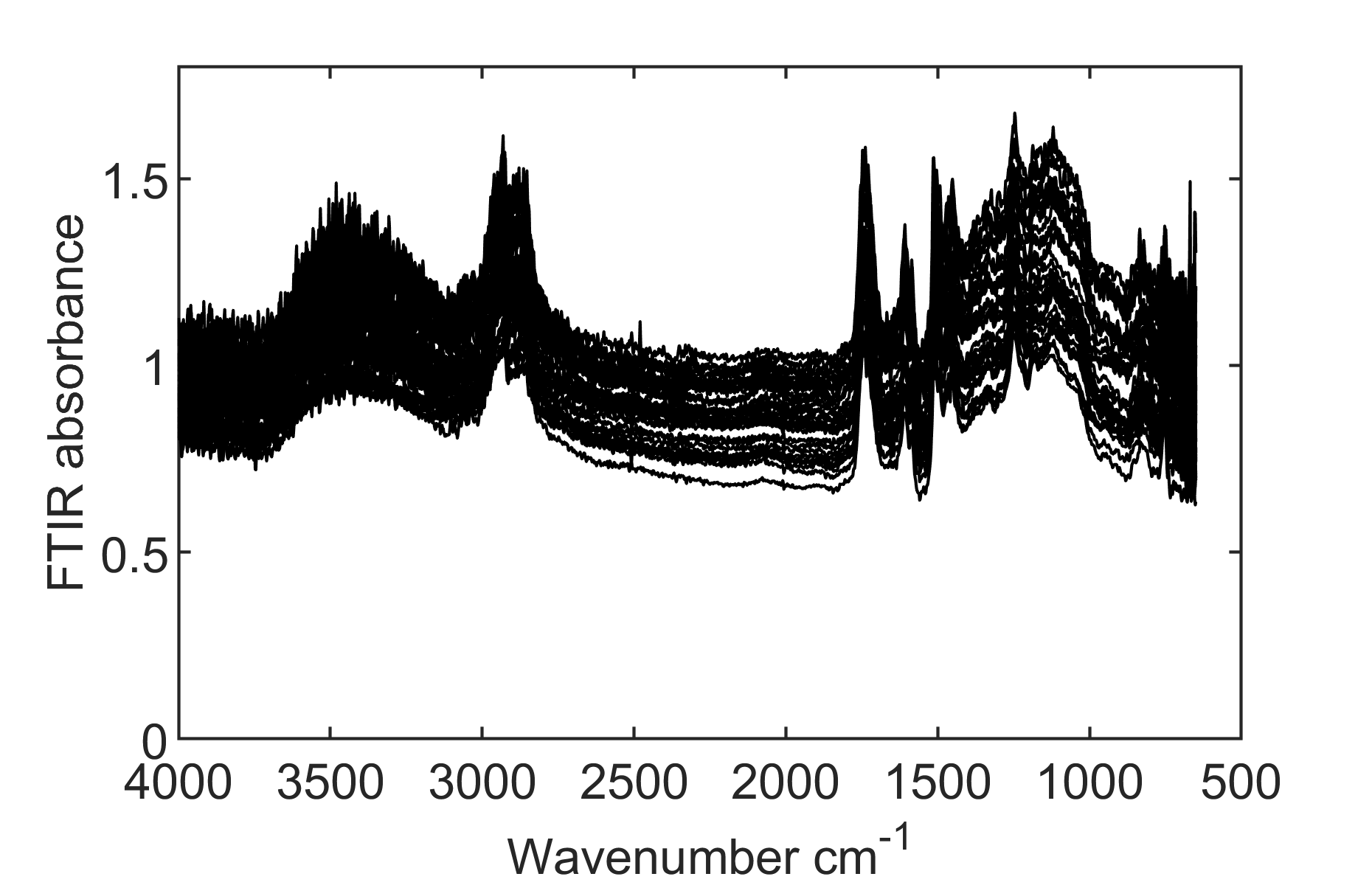}
    \newline
    \subfigure (a) Pre-exposure FTIR signals obtained before plasma exposure
      
    \includegraphics[width=3.5 in]{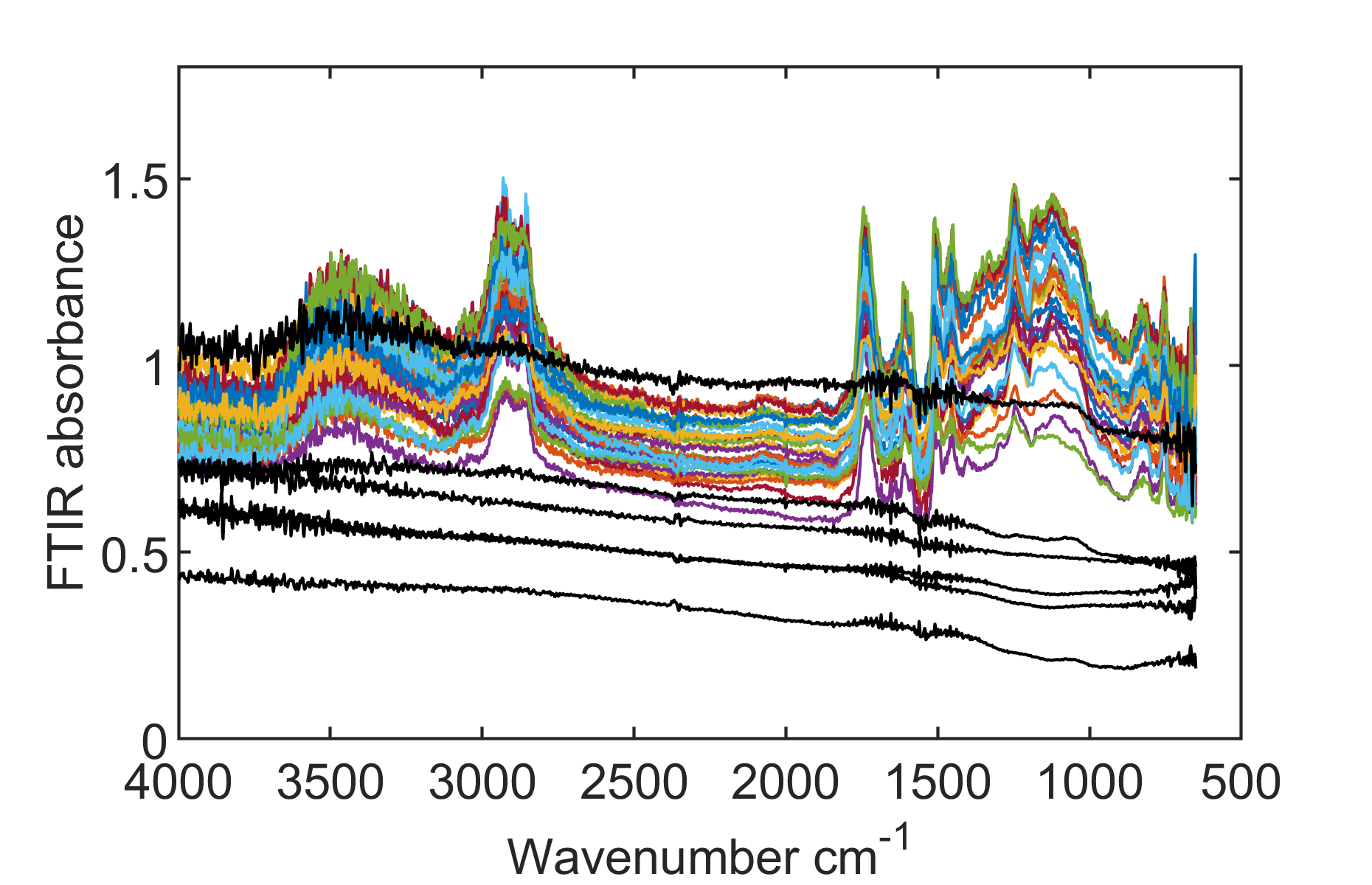}  
    \newline
    \subfigure (b) Post-exposure FTIR signals obtained after plasma exposure at various heights
 
    \caption{Raw FTIR signals collected from CFRP coupons in the experiments shown in Fig.~\ref{fig1}. 
    } 
    \label{fig2}
\end{figure}

To tackle these challenges and achieve our objective, we developed a statistical framework in this study that (i) characterizes the underlying uncertainty of the FTIR spectra, and (ii) describes the nonlinear effect on the post-treatment FTIR spectra. Based on the proposed model, a two-step procedure is developed. In each step, an optimization problem is formulated to estimate the template FTIR signal that is representative for all measurement spectra and to represent the effect of plasma exposure respectively. To validate our methodology, we conducted experiments and collected FTIR spectra measurements from samples before and after the plasma exposure.  These measurements are shown in Fig.~\ref{fig2}.

The contribution of our work is twofold. In terms of statistical analysis, we proposed a preprocessing algorithm to derive a template spectrum from FTIR spectra obtained from repetitive measurements and a general strategy to quantify the nonlinear treatment effect on the spectrum measurements. The methodology is applicable to a large variety of studies that involves understanding the treatment effects on spectral measurements subject to similar multiplicative and offset uncertainty. In terms of manufacturing engineering, we gained an understanding of the effect of plasma exposure on the CFRP material for the first time, and identified several chemical bonds that are affected by plasma exposure.

The remaining part of the article is organized as follows. In Section II, we look into the data characteristics and present the statistical analysis procedure in detail. In Section III, we verify the performance of the proposed two-step algorithm on synthetic data. In section IV, we demonstrate the analytical result on experimental real data. Finally, we conclude this article in Section V.

\section{Uncertainty Analysis and Proposed Framework}

In this section, we first analyze the sources and patterns of the variations of signal data collected from the FTIR metrology, and then propose a statistical model that describes the uncertainty of these signals. Based on the statistical model, we formulate two optimization problems to obtain the template signal and capture the effect of the surface treatment.

\subsection{Uncertainties in FTIR Metrology and Spectra Modeling}

\begin{figure}
    \centering
    \includegraphics[width=3.5 in]{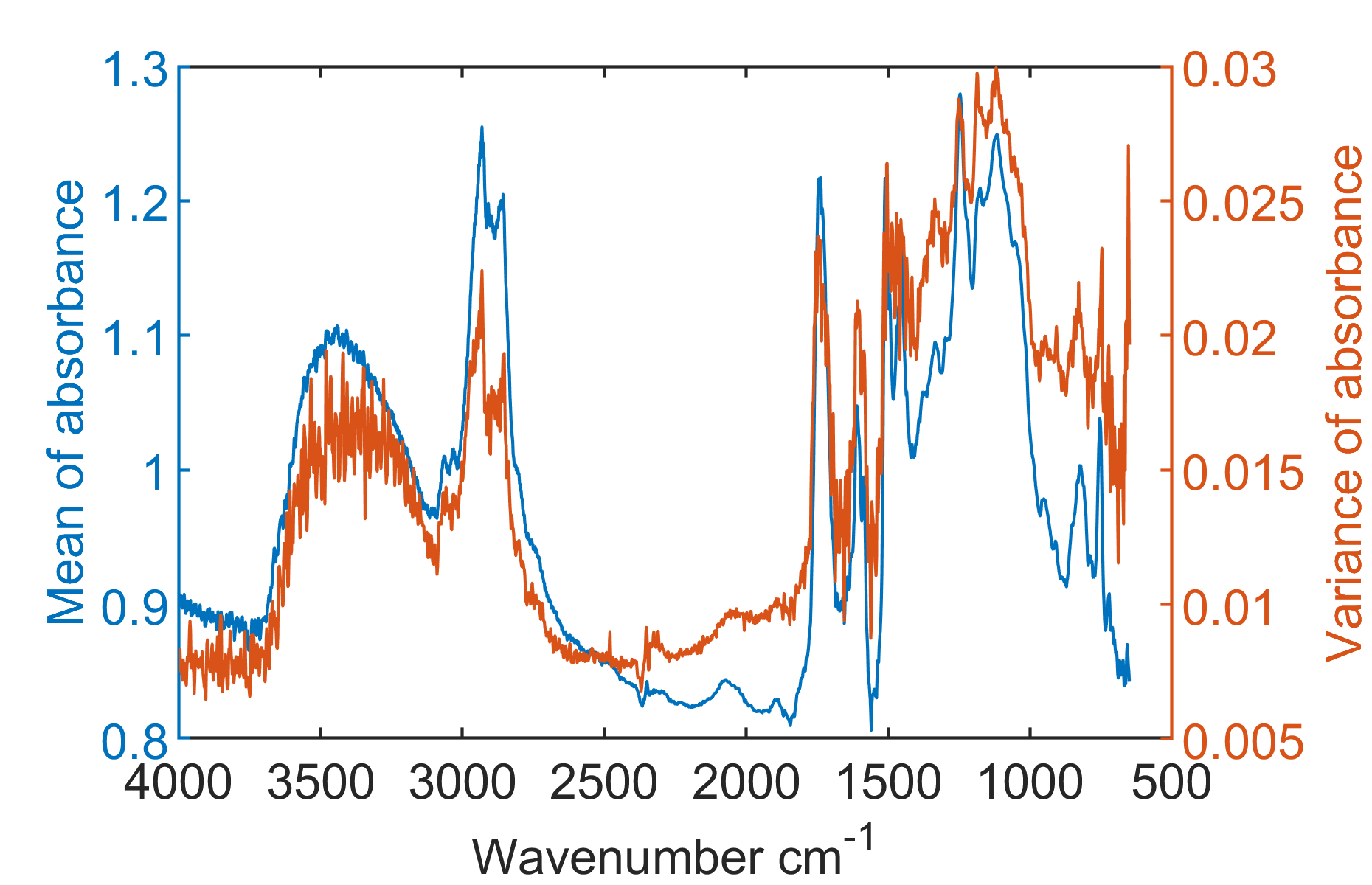}  
    \caption{The sample variance of the FTIR signal (the orange curve) is closely related to the sample mean of the FTIR signal (the blue curve) obtained from the experiment in Fig.~\ref{fig1}.}
    \label{fig:1bnew}
\end{figure}

Recall that the measurements of the FTIR spectrum are subject to offset shift, multiplicative error, and other noises. Among these uncertainties, the offset shift and multiplicative error are dominant. As shown in Fig.~\ref{fig2} (a), signals obtained from the pre-treatment surfaces are similar in shapes, as if generated by vertically stretching and moving a template signal randomly. We refer to the variation of the vertical scale as “multiplicative error”, and the variation of the vertical location as the “offset shift”. According to the literature \cite{afseth2012extended,bassan2011light,cornel2008quantitative},  the offset shift is mainly caused by light scattering and the multiplicative error is mainly caused by optical path length variations related to the hand-held nature of the device. We can also see that the variance of the noise is dependent on the shape of the signal since the noise is more significant at the peaks of the signal, as shown in Fig.~\ref{fig:1bnew}. This phenomenon is common in light-based spectroscopy. For example, Yue et.al \cite{yue2017generalized} summarized multiple error sources in greater detail for Ramen spectra, a similar metrology in material science. 
From this observation, we assume that all pre-treatment signals come from a common template signal denoted by $\mathbf{x}_{0} \in \mathbb{R}^{p}$. Then, the $i$-th pre-treatment signal $\mathbf{x}_{0, i}=\left(x_{0, i}^{(1)}, \ldots, x_{0, i}^{(p)}\right)^{\top} \in \mathbb{R}^{p}$  is modeled as
\begin{equation}
  \mathbf{x}_{0,i}=
     a_{0,i} \left(\mathbf{x}_{0}+ \boldsymbol{\varepsilon}_{0, i}\right)+b_{0, i} \mathbf{1},~ i=1, \ldots, n  
     \label{eq:pretreat}
\end{equation}
Here $a_{0, i} \in \mathbb{R}$  is the factor for the multiplicative error, $b_{0, i} \in \mathbb{R}$  is the offset, and $\boldsymbol{\varepsilon}_{0, i} \in \mathbb{R}^p$  is the noise vector, whose elements are independent and follow  $\varepsilon_{0, i}^{(j)} \sim N\left(0, \sigma^{2}\right),~i=1, \dots, n ;~ j=1, \dots, p$. The vector $\mathbf{1}=(1, \ldots, 1)^{\top} \in \mathbb{R}^{p}$. Note that in the formulation (\ref{eq:pretreat}), the parameters 
are not identifiable without constraints, since the tuple $(a_{0,i},~ \mathbf{x}_{0},~ b_{0,i})$ and $(ka_{0,i},~ (\mathbf{x}_{0}-\mu\mathbf{1})/k,~ b_{0,i}+a_{0,i}\mu)$ correspond to the same probability model for $\mathbf{x}_{0,i}$ for any $k$ and $\mu$. 
To make the model identifiable, we apply two additional constraints on $\mathbf{x}_{0}$: 
\begin{equation}
\left\|\mathbf{x}_{0}\right\|^{2}=\mathbf{x}_{0}^{\top} \mathbf{x}_{0}=1, \quad \mathbf{x}_{0}^{\top} \mathbf{1}=0.
\label{eq:cstr1}    
\end{equation}

In this formulation, we assume that the measurement taken every time are independent with each other. 
This is because the area that the FTIR equipment samples on is very small, and thus the hand-held FTIR equipment cannot obtain pre-treatment or post-treatment measurements at exactly the same location on the sample. Also, note that the random error $\boldsymbol{\varepsilon}_{0,i}$ is firstly added to the template $\mathbf{x}_{0}$, and then affected by the multiplicative error $a_{0,i}$. Therefore, the standard deviation of the noise on the signal is proportional to the multiplicative error, and thus every variable of the observed pre-treatment signal follows a normal distribution with different variance $x_{0, i}^{(j)} \sim N\left(a_{0, i} x_{0}^{(j)}+b_{0, i}, a_{0, i}^{2} \sigma^{2}\right)$. This observation corroborates with the property of the FTIR signal as discussed earlier. 

The proposed model is different from the model in Yue et al. \cite{yue2017generalized}, which proposed   $x_{0,i}^{(j)} \sim N\left(x_{0}^{(j)}, a x_{0}^{(j)}+b\right)$, because the multiplicative error and outfit error are not considered in Yue's model.  
As will be seen later, estimating $\mathbf{x}_0$ in our model is different from simply denoising the raw signals analytically. Also, our representation of the signals is different from the MSC approach \cite{afseth2012extended,cornel2008quantitative}, which implicitly assumes that all signals have the same magnitudes of errors in calculating the sample mean as the template signal. 

The FTIR measurements obtained after the treatment procedure, such as plasma exposure in our motivating example, are also subject to offset shift and multiplicative error. Moreover, the treatment 
effect leads to the shape change from the template signal. The \textit{pattern} of the signal change caused by the treatment effect is assumed 
to be the same across all signals, and are irrelevant of the strength of the treatment, whereas the \textit{magnitude} of the change relates to the strength of the treatment. In the motivating example,  the change of the FTIR signal is of greater magnitude when the plasma nozzle is closer to the CFRP coupon. From the above reasoning, the post-treatment template is modeled as $\mathbf{x}_{0}+\delta_{i} \mathbf{g}$, where   $\mathbf{g} \in \mathbb{R}^{p}$ 
represents the pattern of modification:  the change of the template caused by plasma exposure. The value $\delta_{i}$ represents the magnitude of
modification on the $i$-th spectrum, and $\boldsymbol{\delta}=\left(\delta_{1}, \ldots, \delta_{n}\right)$ is referred to as the vector of effects. Therefore, we assume the following model for the post-treatment error: 	
\begin{equation}
    \mathbf{x}_{1, i}=a_{1, i}  
    \left(\mathbf{x}_{0}+\delta_{i} \mathbf{g}+\boldsymbol{\varepsilon}_{1, i}\right)+b_{1, i} \mathbf{1},~ i=1, \ldots, n
    \label{eq:posttreat}
\end{equation}

Like the pre-treatment model \eqref{eq:pretreat}, $a_{1, i}$ and $b_{1, i}$  represent the multiplicative error and the offset shift respectively, and they are not of interest in our inference. Note that here $a_{1, i}$ and $ b_{1, i}$   are also independent with the  $a_{0, i}$ and  $b_{0, i}$ in the pre-treatment signals. The error $\boldsymbol{\varepsilon}_{1, i}$  is the noise vector following the same distribution as $\boldsymbol{\varepsilon}_{0, i}$, and is independent with  $\boldsymbol{\varepsilon}_{0, i}$. Similar to the pre-treatment model \eqref{eq:pretreat},  the parameters in the post-treatment model  \eqref{eq:posttreat} are  not  identifiable, and therefore, the additional constraint $\| \mathbf{g}\|^2= 1$ is imposed. Also, we encourage $\mathbf{g}$ to be close to 0 at most elements, since the surface treatment usually triggers chemical reactions among specific chemical compositions rather than for all the chemical compositions on the material surface.
Note that we provide a very flexible model for $\boldsymbol\delta$, the magnitude vector.

 In the above pre-treatment and post-treatment models, the effect of the exterior treatment on the surface and the uncertainty involved with the FTIR measurement are fully characterized. In the following subsections, we develop a two-step procedure, to  estimate the template signal $\mathbf{x}_{0}$, the pattern of the signal change $\mathbf{g}$ and the magnitude of modification $\delta$. In the first step, we use the pre-treatment signals to estimate the template signal $\mathbf{x_0}$, and in the second step, we regard the template signal estimated in Step 1 as known and fixed and use post-treatment signal to estimate the
pattern of modification $\mathbf{g}$ and the vector of effect $\boldsymbol{\delta}$. The detailed procedure are introduced in the following subsections. 

\subsection{Step 1: Estimate the Template Spectra  $\mathbf{x}_{0}$}
In Step 1, our objective is to estimate the template signal $\mathbf{x}_{0}$ in model \eqref{eq:pretreat} from the pre-treatment FTIR measurements $\mathbf{x}_{0,1}, \mathbf{x}_{0,2}, \ldots, \mathbf{x}_{0, n},$ In Cornel \cite{cornel2008quantitative}, some ad-hoc methods have been provided to estimate the template signal, such as through standardizing and averaging the measurements signals. However, these methods have no guarantee to eliminate the error to the greatest extent. 
In this study, we propose to use the maximmum likelihood estimation (MLE) principle to develop a computational efficient algorithm to estimate $\mathbf{a}_{0},~ \mathbf{b}_{0}$, and $\mathbf{x}_{0}$.

Based on the pre-treatment model (1), it is clear that $\mathbf{x}_{0, i} \sim N\left(a_{0, i} \mathbf{x}_{0}+b_{0, i} \mathbf{1},~ a_{0, i}^{2} \sigma^{2}\mathbf{I}\right)$ , where $\sigma^2 = \mathrm{var}(\varepsilon_{0, i}^{(j)})$. 
Thus, the log-likelihood function is 
\begin{multline}
\ell\left(\boldsymbol{a}_{0}, \boldsymbol{b}_{0}, \mathbf{x}_{0} ; \mathbf{x}_{0, \mathrm{i}}\right)=   
-\frac{n}{2} \ln \left(2 \pi \sigma^{2}\right) \\
-\sum_{i=1}^n\frac{\lVert\mathbf{x}_{0, i}-a_{0, i} \mathbf{x}_{0}-b_{0 i,} \mathbf{1}\rVert^2}{2 a_{0, i}^{2} \sigma^{2}}
-\sum_{i=1}^{n} \ln \left|a_{0, i}\right|
\label{MLE}
\end{multline}

Note that we need to optimize the log-likelihood function subject to the constraints in \eqref{eq:cstr1}. It is computationally challenging to find the solution for this constrained optimization jointly over the parameter space ($\mathbf{a}_0,~ \mathbf{b}_0,~ \sigma^2,~\mathbf{x}_0$) of the dimension $2n+1+(p-1) = 2n+p$. 
To circumvent this difficulty and focus on the parameters $\mathbf{x}_0$, we propose to investigate the main term related to $\mathbf{x}_0$ in the \eqref{MLE} and minimize the objective function:
\[ \sum_{i=1}^{n}\lVert\mathbf{x}_{0, i}-a_{0, i} \mathbf{x}_{0}-b_{0,i} \mathbf{1}\rVert^{2} /a_{0, i}^{2} \]

Re-parameterize  $c_{0, i}=a_{0, i}^{-1}$ and $d_{0, i}=-a_{0, i}^{-1} b_{0, i}$ and apply the constraints in the model (1), the following optimization problem needs to be solved
 \begin{align}
     \min _{\mathbf{x}_{0}, \mathbf{c}_{0}, \mathbf{d}_{0}} \sum_{i=1}^{n} & \left\|c_{0, i}\mathbf{x}_{0, i}+d_{0, i} \mathbf{1}-\mathbf{x}_{0}\right\|^{2} \notag \\
     \text {subject to} & \quad \mathbf{x}_{0}^{\top} 1=0,~\left\|\mathbf{x}_{0}\right\|^{2}=1 \label{eq:step1}
 \end{align}
where $\mathbf{c}_{0}=\left(c_{0,1}, \ldots, c_{0, n}\right)^{\top}$ and $\mathbf{d}_{0}=\left(d_{0,1}, \ldots, d_{0, n}\right)^{\top}$, and
$\|\cdot\|$ represents the Euclidian norm. The parameters $a_{0, i},~ b_{0, i}$ can be computed by $c_{0, i}^{-1}$ and $-c_{0, i}^{-1} d_{0, i},$ respectively. 



Given $\mathbf{x}_{0}$, the optimal value of $c_{0, i},~ d_{0, i}$  can be calculated from solving least square problems: 
 \[\left[ \begin{array}{c}{\hat{c}_{0, i}\left(\mathbf{x}_{0}\right)} \\ {\hat{d}_{0, i}\left(\mathbf{x}_{0}\right)}\end{array}\right]=\left[ \begin{array}{cc}{\mathbf{x}_{0, i}^{\top} \mathbf{x}_{0, i}} & {\mathbf{x}_{0, i}^{\top} \mathbf{1}} \\ {\mathbf{1}^{\top} \mathbf{x}_{0, i}} & {p}\end{array}\right]^{-1} \left[ \begin{array}{c}{\mathbf{x}_{0, i}^{\top} \mathbf{x}_{0}} \\ {\mathbf{1}^{\top} \mathbf{x}_{0}}\end{array}\right]\]

Plug them in \eqref{eq:pretreat} and denote $\mathbf{H}_{i}=\mathbf{x}_{0, i} \mathbf{x}_{0, i}^{\top}\left(\mathbf{x}_{0, i}^{\top} \mathbf{x}_{0, i}\right)^{-1}\left[\mathbf{I}-\frac{\mathbf{11^{\top}}}{p}\right]$, the objective of (3) is transformed to
\[f\left(\mathbf{x}_{0}, \mathbf{c}_{0}\left(\mathbf{x}_{0}\right), \mathbf{d}_{0}\left(\mathbf{x}_{0}\right)\right)=\sum_{i=1}^{n}\left\|\mathbf{H}_{i} \mathbf{x}_{0}-\mathbf{x}_{0}\right\|^{2}\]

Then, it can be further written as $f\left(\mathbf{x}_{0}, \mathbf{c}_{0}\left(\mathbf{x}_{0}\right), \mathbf{d}_{0}\left(\mathbf{x}_{0}\right)\right)=\mathbf{x}_{0}^{\top} \mathbf{M} \mathbf{x}_{0}$, where $\mathbf{M}=\sum_{i=1}^{n} \mathbf{V}_{i}^{\top} \mathbf{V}_{i}$, and $\mathbf{V}_{i}=\mathbf{H}_{i}-\mathbf{I}$, and thus the problem \eqref{eq:step1} is transformed to 
 \begin{align}
     \min _{\mathbf{x}_{0}, \mathbf{c}_{0}, \mathbf{d}_{0}} & \mathbf{x}_{0}^{\top} \mathbf{M} \mathbf{x}_{0} \notag \\
     \text {subject to} \quad  & \mathbf{x}_{0}^{\top} \mathbf{1}=0  \text ,~\left\|\mathbf{x}_{0}\right\|^2 =1  \label{eq:eigenprob}
 \end{align}

 The objective function becomes a quadratic function of $\mathbf{x}_{0}$, and thus the problem is essentially an eigen problem with linear constraint. It can be shown \cite{golub1973some} that the solution $  \hat{\mathbf{x}}_{0} $ to this problem is $\mathbf{P v}$, where  $\mathbf{P}=\mathbf{I}-\mathbf{1 1}^{\top}$ and $ \mathbf{v}$ is the eigen vector of $ \mathbf{P M P}$ corresponding to the smallest  eigen value.

\subsection{Step 2: Estimate $\boldsymbol{\delta}$  and $\mathbf{g}$ from the Post-treatment Spectra }
In Step 1, the estimation of the template signal $\mathbf{x}_{0}$ was obtained from the last step based on pre-treatment signals. In Step 2, this template is regarded as known and our objective is to estimate $\mathbf{g} \in \mathbb{R}^{p}$ and $\boldsymbol{\delta}=\left(\delta_{1}, \ldots, \delta_{n}\right)^{\top}$  from the post-treatment signals. Like problem \eqref{eq:step1}, we can find the solution of $\boldsymbol{\delta},~ \mathbf{g}$  from the following least square problem (\ref{eq:step2}): 
\begin{align}
    \min_{\boldsymbol{\delta}, \boldsymbol{g}_{1},\mathbf{c}_{1}, \mathbf{d}_{1}}  \sum_{i=1}^{n}&\left\|c_{1, i} \mathbf{x}_{1, i}+d_{1, i} \mathbf{1}-\left(\mathbf{x}_{0}+\delta_{\mathrm{i}} \mathbf{g}\right)\right\|_{2}^{2} \notag \\
    \text{subject to} & \quad \|\mathbf{g}\|^2=1 \label{eq:step2}
\end{align}

However, similar to the problem in Step 1, the parameters $c_{1,i}$, $d_{1,i}$, and $\delta_i$ are not identifiable without further constraints.
To solve the identifiability issue, the pattern of modification $\mathbf{g}$ can be any function obtained from the linear combination of $\mathbf{\tilde{g}, 1}$ and $\mathbf{x}_{0}$, where $\mathbf{\tilde{g}}$ is the component of $\mathbf{g}$ in the null space of $\mathbf{1}$ and $\mathbf{x}_{0}$. In what follows, we first find the vector $\tilde{\mathbf{g}}$ by solving problem \eqref{eq:step2} in addition to two constraints
on $\mathbf{g}: \mathbf{g}^{\top} \mathbf{x}_{0}=0$ and $\mathbf{g}^{\top} \mathbf{1}=0,$ which leads to problem \eqref{eq:step2constr}: 
\begin{align}
    \min _{\boldsymbol{\delta}, \mathbf{\tilde{g}},\mathbf{c}_{1}, \mathbf{d}_{1}} \sum_{i=1}^{n}&\left\|c_{1, \mathrm{i}}  \mathbf{x}_{1, i} +d_{1, i} \mathbf{1}-\left(\mathbf{x}_{0}+\delta_{i} \tilde{\mathbf{g}}\right)\right\|_{2}^{2} \notag \\
    \text{subject to} & \quad \tilde{\mathbf{g}}^{\top} \mathbf{x}_{0}=\mathbf{0},~ \tilde{\mathbf{g}}^{\top} \mathbf{1}=\mathbf{0} \label{eq:step2constr}
\end{align}

After that, we discuss how to find the pattern of modification $\mathbf{g}=\epsilon_{1} \mathbf{x}_{0}+\epsilon_{2} \mathbf{1}+\sqrt{1-\epsilon_{1}^{2}-\epsilon_{2}^{2}} \tilde{\mathbf{g}}$  with the best interpretability. 

\subsection{Solution Procedure for Problem \eqref{eq:step2constr}}
For the simplicity of notations, we drop the subscript $“1”$ in
$c_{1, i}$ and $d_{1, i}$ and $\mathbf{x}_{1, i},$ and drop the tilde from $\tilde{\mathbf{g}}$.
Then, the objective function in \eqref{eq:step2constr} is denoted by 
\begin{align}
    F(\boldsymbol{\delta}, \mathbf{g}, & \mathbf{c}, \mathbf{d}) = \sum_{i=1}^{n} \sum_{j=1}^{p}\left(c_{i} x_{i j}+d_{i}-x_{0, j}-\delta_{i} g_{j}\right)^{2}  \notag \\
    =& \left\|\left(\mathbf{c} \mathbf{1}^{\top}\right) \odot \mathbf{X}+\mathbf{d} \mathbf{1}^{\top}-\mathbf{1} \mathbf{x}_{0}^{\top} -\boldsymbol{\delta} \mathbf{g}^{\top}\right\|_{F}^{2} \label{eq:step2F}
\end{align}
where $\mathbf{c}=\left(c_{1}, c_{2}, \ldots, c_{ n}\right)^{\top}, \mathbf{d}=\left(d_{1}, d_{2}, \ldots, d_{n}\right)^{\top}$, and  $\mathbf{X}=\left[\mathbf{x}_{1}, \ldots, \mathbf{x}_{ n}\right]^{\top}$. The operator $\odot$ is the
elementwise product of two matrix. We solve the problem \eqref{eq:step2F} with the following block-wise coordinate descent method: 

\begin{algorithm}[htb]
  \caption{ Block-wise Coordinate Descent Algorithm}
  \label{alg.Framwork}
  \begin{algorithmic}[1]
\State Initialization: $\boldsymbol{\delta}\gets \mathbf{0}$ and arbitrary $\mathbf{g}$

\Loop
     \State Given $\boldsymbol{\delta}$ and $\mathbf{g}$, update $\mathbf{c}$ and $\mathbf{d}$:
     
     $\left[\mathbf{c}, \mathbf{d}\right]\gets\arg \min _{\mathbf{c}, \mathbf{d}} F\left(\boldsymbol{\delta}, \mathbf{g}, \mathbf{c}, \mathbf{d}\right)$;
     
     \State Given $\mathbf{c}$ and $\mathbf{d}$, update $\boldsymbol{\delta}$ and $\mathbf{g}$:

         $\left[\boldsymbol{\delta}, \mathbf{g}\right]\gets\arg \min _{\boldsymbol{\delta}, \mathbf{g}} F\left(\boldsymbol{\delta}, \mathbf{g}, \mathbf{c}, \mathbf{d}\right)$
        
         subject to $\mathbf{g}^{\top} \mathbf{1}=0;~ \mathbf{g}^{\top} \mathbf{x}_{0}=0$ and $\|\mathbf{g}\|^2=1$;

\EndLoop
  \end{algorithmic}
\end{algorithm}

     
  


In this algorithm, the optimization problem in line 3 can be solved by multiple least-square problems to obtain $\left(c_{ i}, d_{ i}\right)^{\top}$, which can be seen from (8). The closed form solution to line 3 is 
\[\left[ \begin{array}{c}{c_{i}} \\ {d_{i}}\end{array}\right]=\left[ \begin{array}{cc}{\mathbf{x}_{i}^{\top} \mathbf{x}_{i}} & {\mathbf{x}_{i}^{\top} \mathbf{1}} \\ {\mathbf{1}^{\top} \mathbf{x}_{i}} & {p}\end{array}\right]^{-1} \left[ \begin{array}{c}{\mathbf{x}_{i}^{\top}\left(\mathbf{x}_{0}+\delta_{i} \mathbf{g}\right)} \\ {\mathbf{1}^{\top}\left(\mathbf{x}_{0}+\delta_{i} \mathbf{g}\right)}\end{array}\right]\]

Denote $\mathbf{M}=\left(\mathbf{c} \mathbf{1}^{\top}\right) \odot \mathbf{X}+\boldsymbol{d} \mathbf{1}^{\top}-\mathbf{1} \mathbf{x}_{0}^{\top}$. Line 4 is equivalent with solving the problem \eqref{eq:step2svd}: 
\begin{align}
\left[\boldsymbol{\delta}, \mathbf{g}\right]=\operatorname{arg min}_{\boldsymbol{\delta}, \mathbf{g}}\left\|\mathbf{M}-\boldsymbol{\delta} \mathbf{g}^{\top}\right\|_{F}^{2} \notag \\
\text{subject to} ~ \mathbf{g}^{\top} \mathbf{1}=0,~ \mathbf{g}^{\top} \mathbf{x}_{0}=0 ,~ \left\|\mathbf{g}\right\|^2=1 \label{eq:step2svd}
\end{align}

Note that under the constraint that $\mathbf{g}^{\top} \mathbf{1}=0$ and
$\mathbf{g}^{\top} \mathbf{x}_{0}=0,$ the objective of \eqref{eq:step2svd} can be decomposed to
\begin{align}
&{\left\|\mathbf{M}-\boldsymbol{\delta} \mathbf{g}^{\top}\right\|_{F}^{2}} \notag \\
=&\left\|\mathbf{M}-\widetilde{\mathbf{M}}+\widetilde{\mathbf{M}}-\boldsymbol{\delta} \mathbf{g}^{\top}\right\|_{F}^{2} \notag \\
=&\left\|\widetilde{\mathbf{M}}\right\|_{F}^{2}+\left\|\mathbf{M}-\widetilde{\mathbf{M}}-\boldsymbol{\delta} \mathbf{g}^{\top}\right\|_{F}^{2} \notag
\end{align}
where $\widetilde{\mathbf{M}}=\mathbf{H} \mathbf{M}, \mathbf{H}=\widetilde{\mathbf{X}}_{0}\left(\widetilde{\mathbf{X}}_{0}^{\top} \widetilde{\mathbf{X}}_{0}\right)^{-1} \widetilde{\mathbf{X}}_{0}^{\top},$ and $\widetilde{\mathbf{X}}_{0}=\left[ {\mathbf{1}},~ {\mathbf{x}_{0}}\right]$.
Geometrically,  $\widetilde{\mathbf{M}}$ is obtained by projecting every column of  $\mathbf{M}$ onto the space spanned by $\mathbf{1}$  and $\mathbf{x}_{0}$. Then, the minimizer of problem \eqref{eq:step2svd} can be obtained from solving the reduced rank problem: 
\[\left[\boldsymbol{\delta}, \mathbf{g}\right]=\operatorname{arg min}_{\boldsymbol{\delta}, \mathbf{g}}\left\|\mathbf{M}-\widetilde{\mathbf{M}}-\boldsymbol{\delta} \mathbf{g}^{\top}\right\|_{F}^{2}\]

Let the singular value decomposition of $\mathbf{M}-\widetilde{\mathbf{M}}$ be
$\mathbf{M}-\widetilde{\mathbf{M}}=\sum_{k} \lambda_{k} \mathbf{u}_{k} \mathbf{v}_{k}^{\top},$ with $\lambda_{1} \geq \lambda_{2} \geq \cdots$. The solution $\mathbf{g}$ is the first right singular vector $\mathbf{v}_{1},$ and $\boldsymbol{\delta}$ is $\lambda_{1} \mathbf{u}_{1}$. See,
for example, Theorem 2.4.8 of Golub et al. \cite{golub2012matrix}.

\subsection{Find the Most Interpretable Pattern of Modification $\mathbf{g}$}
After the pattern of modification $\tilde{\mathbf{g}}$  is obtained, the pattern of modification can be any function obtained from the linear combination of $\tilde{\mathbf{g}}, \mathbf{1}$ and $\mathbf{x}_{0}$. As our objective is to understand the change of chemical components, it is desirable that $\mathbf{g}$ be close to zero in most elements. For this reason, we aim to find $\theta \in[0,2 \pi]$ and $\phi \in [0, \pi]$ to minimize
$G(\theta, \phi)=\|\mathbf{g}(\theta, \phi)\|_{1}$, where $\mathbf{g}(\theta, \phi)=\tilde{\mathbf{g}} \cos \phi+\frac{\mathbf{1}}{\sqrt{p}} \cos \theta \sin \phi+\mathbf{x}_{0} \sin \theta \sin \phi,$ Because $\tilde{\mathbf{g}},~ \frac{\mathbf{1}}{\sqrt{p}}$ and $\mathbf{x}_{0}$ are orthogonal with each other and all with
Euclidian norm $1,~\|\mathbf{g}(\theta, \phi)\|_{2}=1$ for all $\theta$ and $\phi$. After solving
\[\left(\theta^{*}, \phi^{*}\right)=\operatorname{arg~min}_{\theta, \phi} G(\theta, \phi),\] 
the pattern of modification is obtained as
\[ \mathbf{g}\left(\theta^{*}, \phi^{*}\right)=\tilde{\mathbf{g}} \cos \phi^{*}+\frac{\mathbf{1}}{\sqrt{p}} \cos \theta^{*} \sin \phi^{*}+\mathbf{x}_{0} \sin \theta^{*} \sin \phi^{*}.\]
 As discussed earlier, the pattern of modification will help us understand the change of chemicals: it provides a map of the treatment effect on all frequency bands, giving comprehensive information on how the chemical bonds change as a result of the surface treatment. 

\section{Simulation studies}
\begin{figure}
    \centering
    \includegraphics[width=3.5 in]{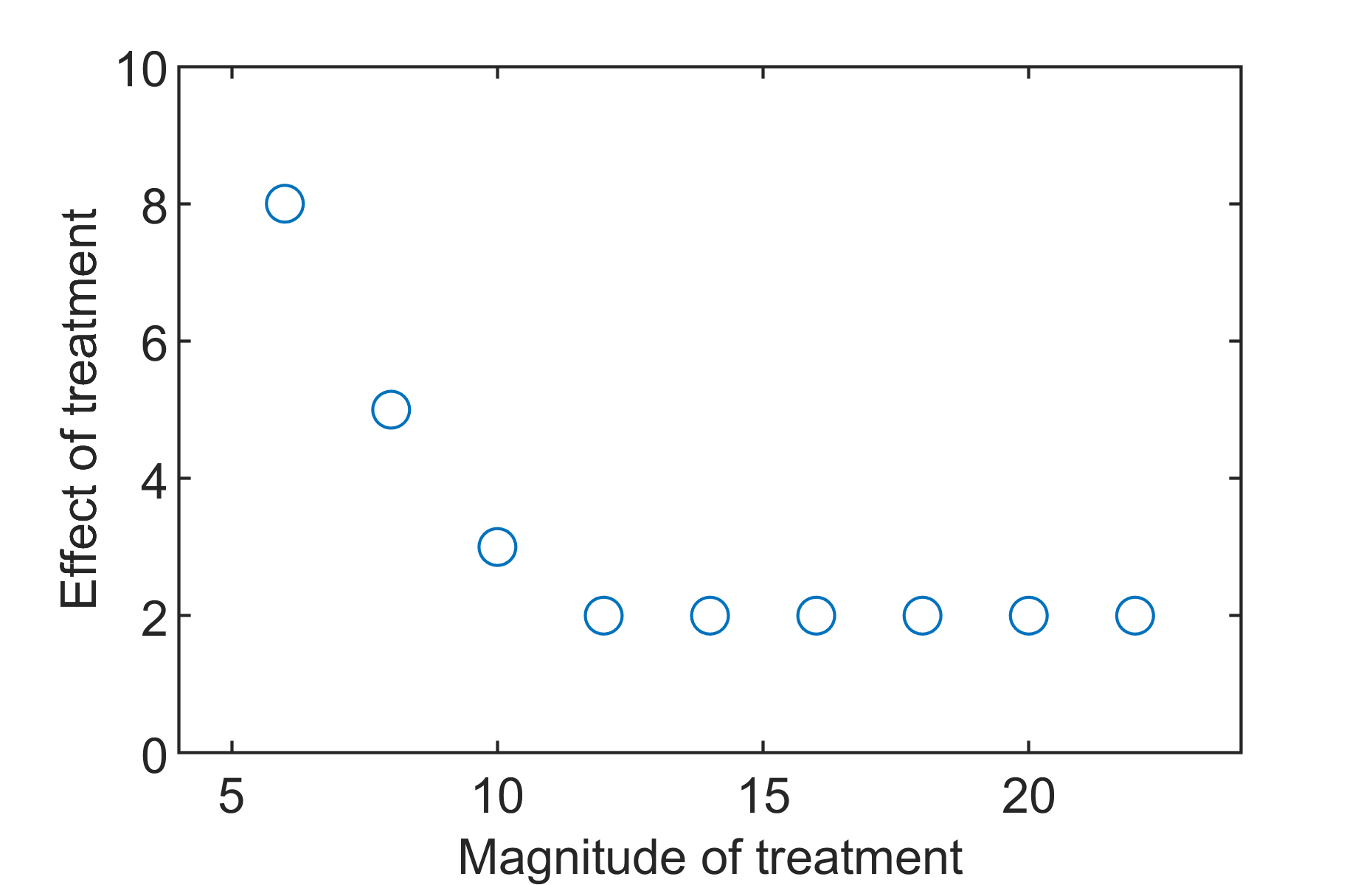}  
    \newline
    \caption{Simulated $\boldsymbol{\delta}$  function.}
    \label{fig5}
\end{figure}

\begin{figure}
    \centering
    \includegraphics[width=3.5 in]{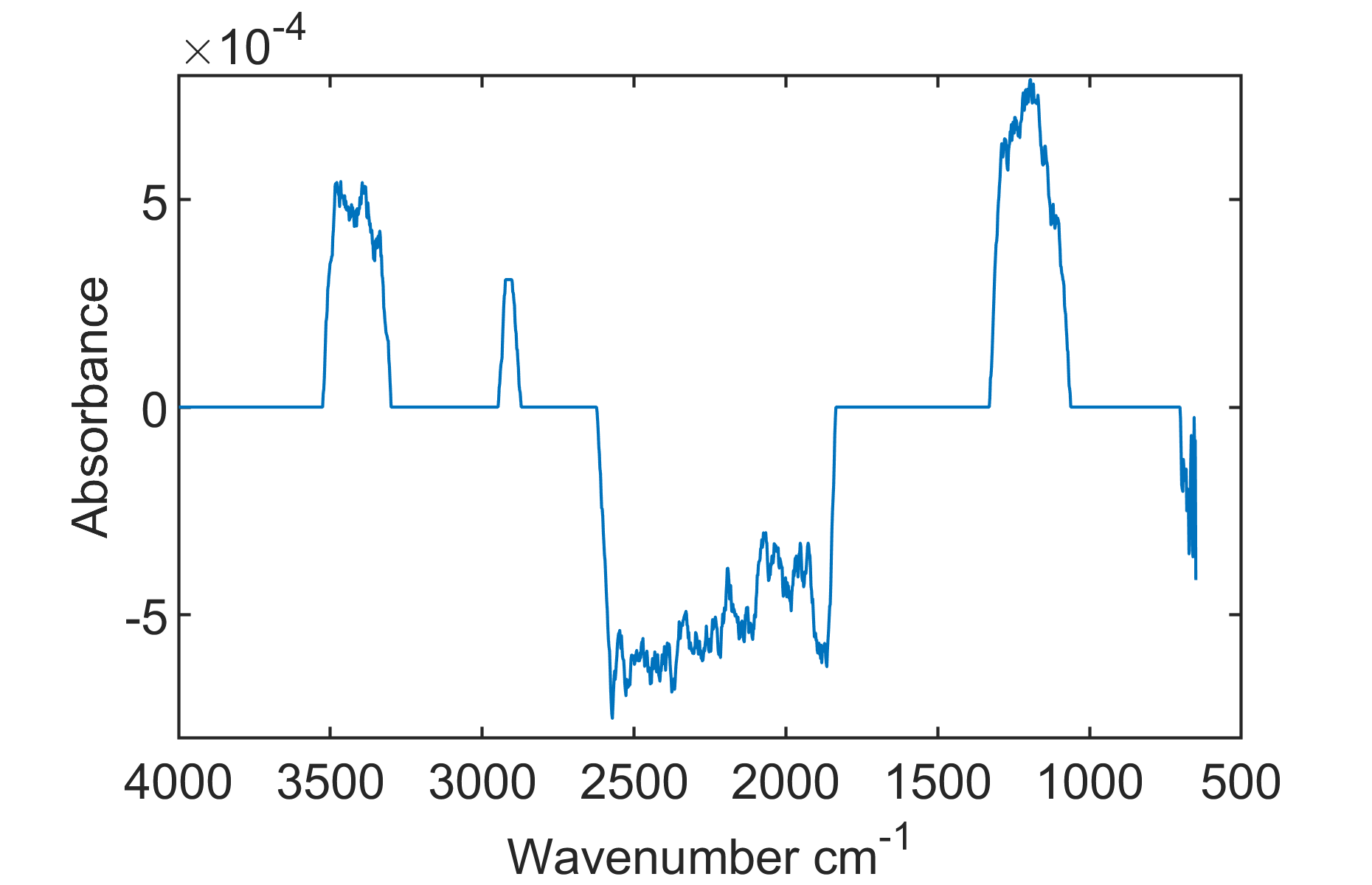}  
    \newline
    \caption{Simulated $\mathbf{g}$  function.}
    \label{fig6}
\end{figure}

\begin{figure}
    \centering
  
    \includegraphics[width=3.5 in]{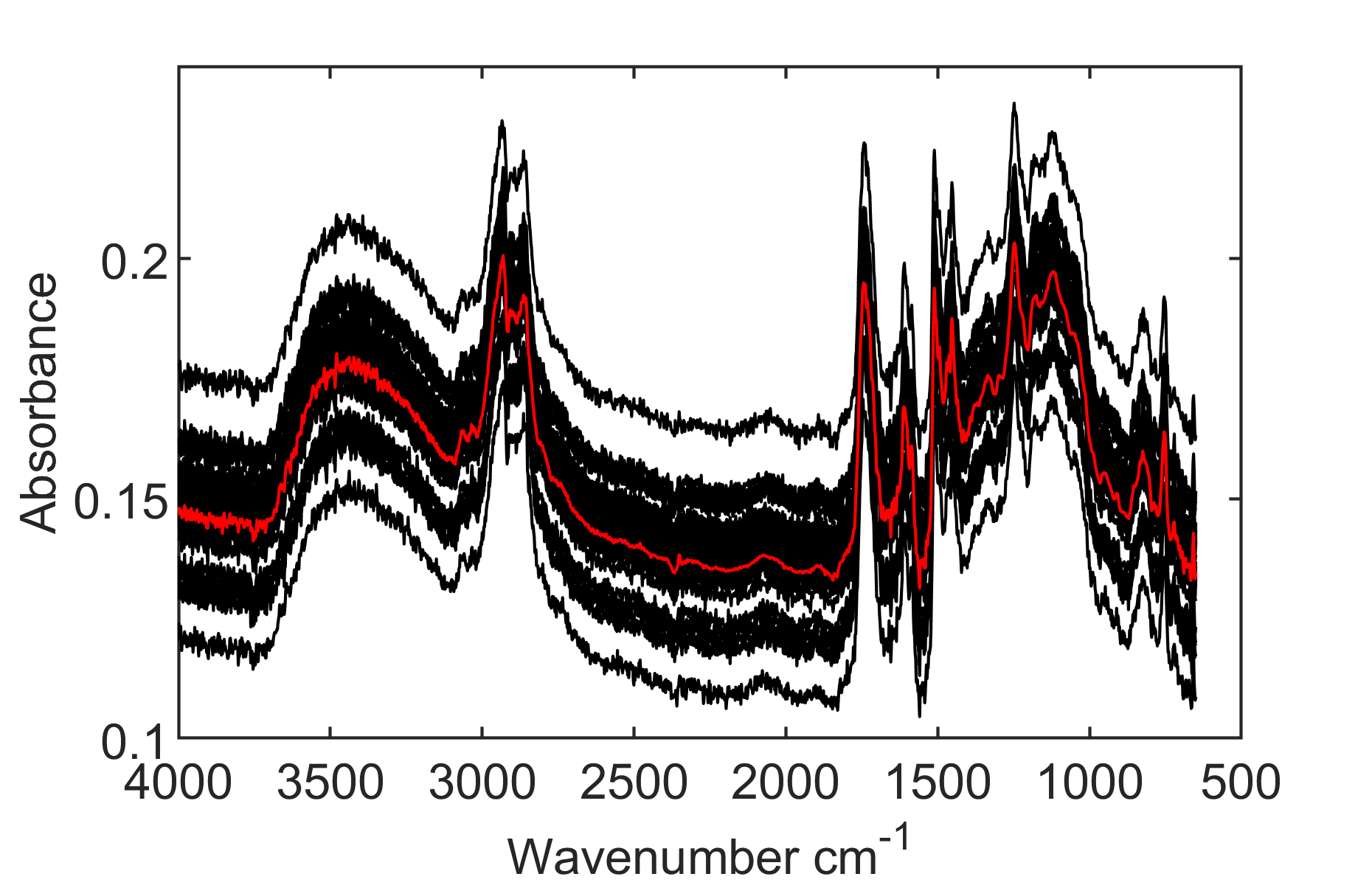}
    \newline
    \subfigure (a) Simulated pre-treatment signals (black curves) and the true template signal $\mathbf{x}_0$ (red curve) 
      
    \includegraphics[width=3.5 in]{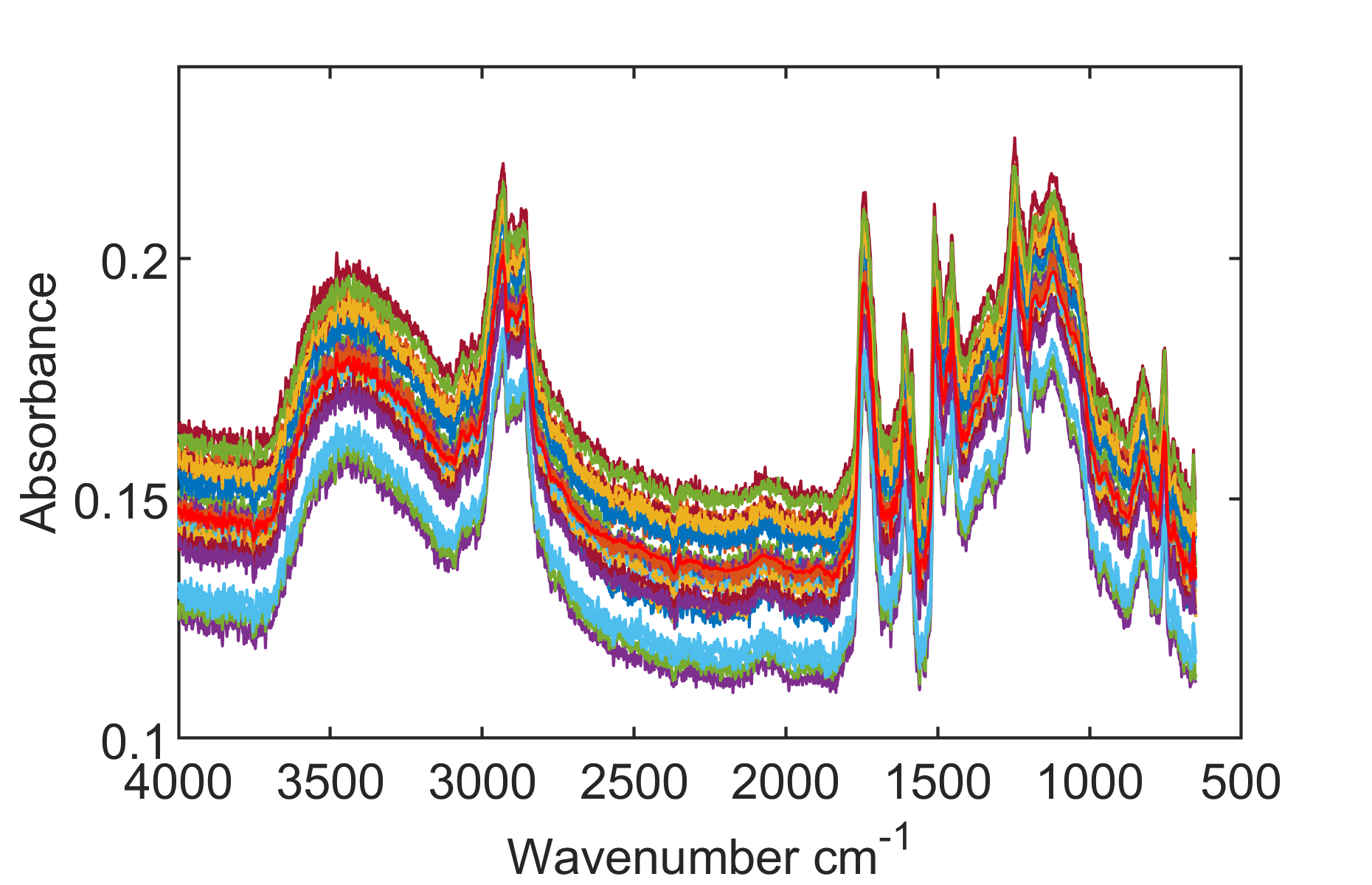}  
    \newline
 \subfigure (b)Simulated post-treatment signals with the same kind of uncertainties
 
    \caption{Simulated FTIR signals.}
    \label{fig3}
\end{figure}

In this section, we implement the two-step algorithm proposed in the last section on synthetic data with the uncertainties described in Section II to validate the effectiveness of the proposed method. In our simulation, the template signal $\mathbf{x}_{0}$ is obtained by averaging and standardizing some pre-treatment signals obtained from the experiment.  
We first generate 33 pre-treatment signals based on the template signal, subject to the random offset, multiplicative error and independent noise. 
To generate the post-exposure signals, we select true value for the vector of effect as $\boldsymbol{\delta}=[8~5~3~2~2~2~2~2~2]^{\top}$, as shown in Fig.~\ref{fig5}. The simulated pattern of modification $\mathbf{g}$ is shown in Fig.~\ref{fig6}. For each $\delta_i$ for $i=1,\ldots,9$, we generated three post-treatment signals, by adding $\delta _i\mathbf{g}$ onto the baseline signals that are also subject to random offset, multiplicative error and independent noise. As a result, we obtain 27 post-exposure signals as shown in Fig.~\ref{fig3}. 
Here, the number of pre-treatment and post-treatment signals are set in consistency with the sample size in our experiments, as detailed in the next section. The analysis of the real data will be illustrated in the next section as well. 

\begin{figure}
    \centering
    \includegraphics[width=3.5 in]{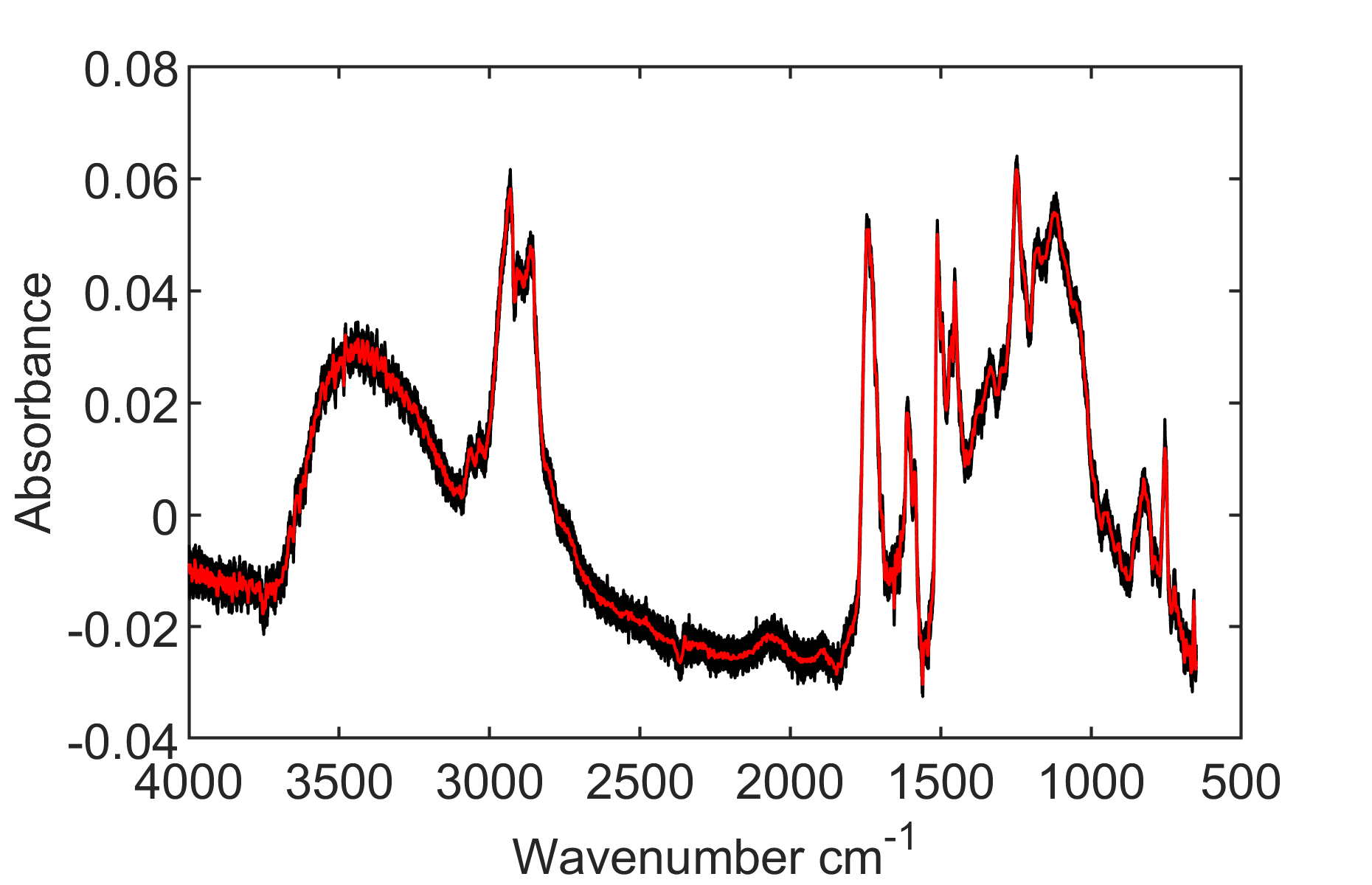}
    \newline
    \subfigure (a) Corrected simulated pre-treatment FTIR signals and the template signal (red curve). 
    
    \includegraphics[width=3.5 in]{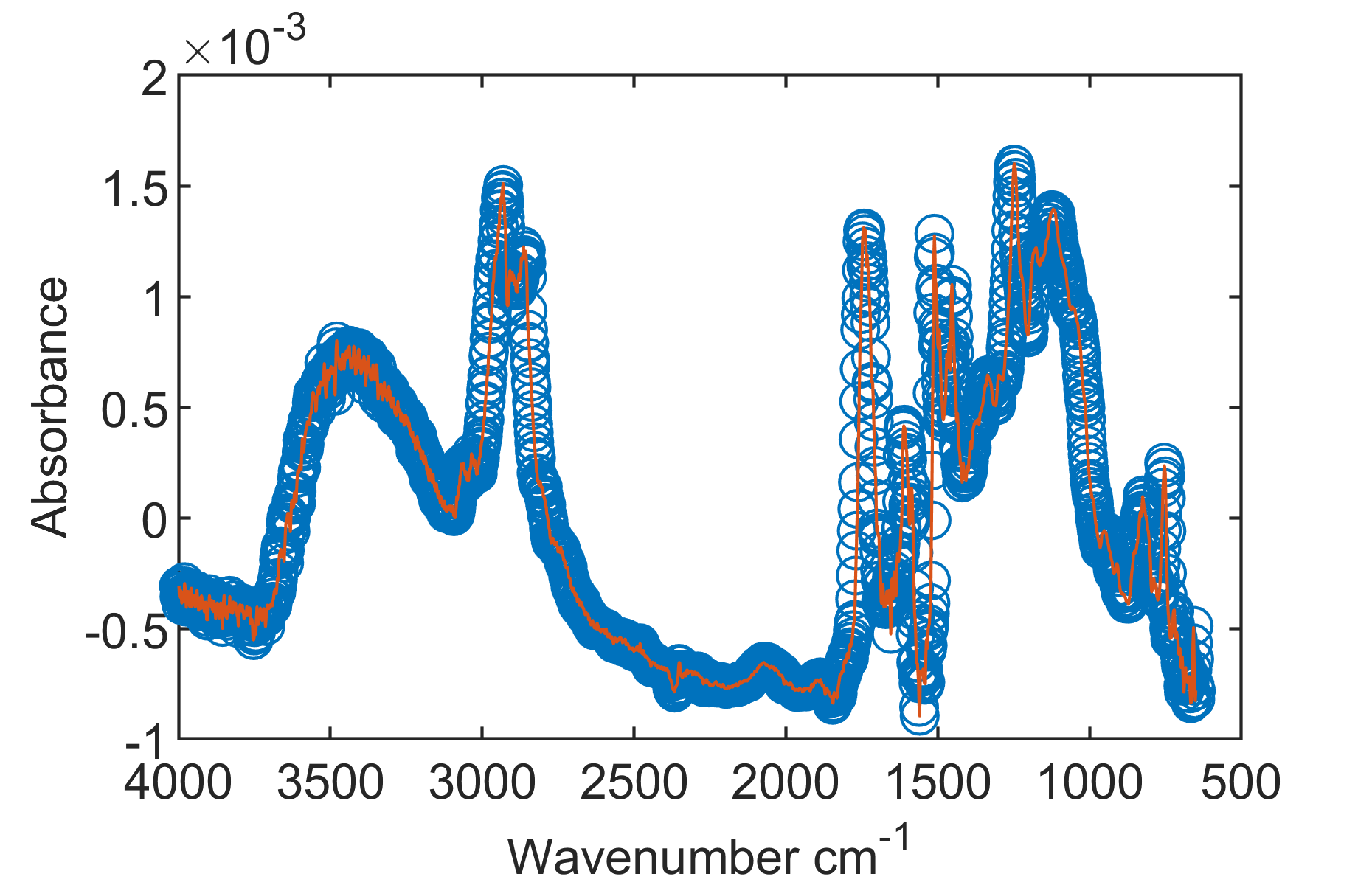}  
    \newline
 \subfigure (b) Comparison of normalized true template signal (red curve) and reconstructed template signal (blue circles) estimated by the proposed algorithm
     \caption{ Estimation results of proposed Step 1 on the simulated data in Fig~\ref{fig3} (a).}
    \label{fig4}
\end{figure}

In Step 1, we estimate the template signal $\mathbf{x}_{0}$  with the aligned pre-treatment FTIR signals by solving problem \eqref{eq:step1}, and the result is shown in Fig.~\ref{fig4} (a). The black curves are the aligned pre-treatment signals $\hat{c}_{0, i} \mathbf{x}_{0, i}+\hat{d}_{0, i} \mathbf{1}$  and the red curve is the template signal $\mathbf{x}_{0}$. Compared with the raw data seen in Fig.~\ref{fig3}, the pre-treatment signals are well aligned. Fig.~\ref{fig4} (b) shows the comparison of the true template signal (red curve) and the reconstructed template signal (blue circles) estimated by the proposed algorithm. The two curves almost coincide, which implies that the template signal is extracted accurately and thus we verify the effectiveness of the proposed algorithm on FTIR spectra pre-treatment. 
\begin{figure}
    \centering
    \includegraphics[width=3.5 in]{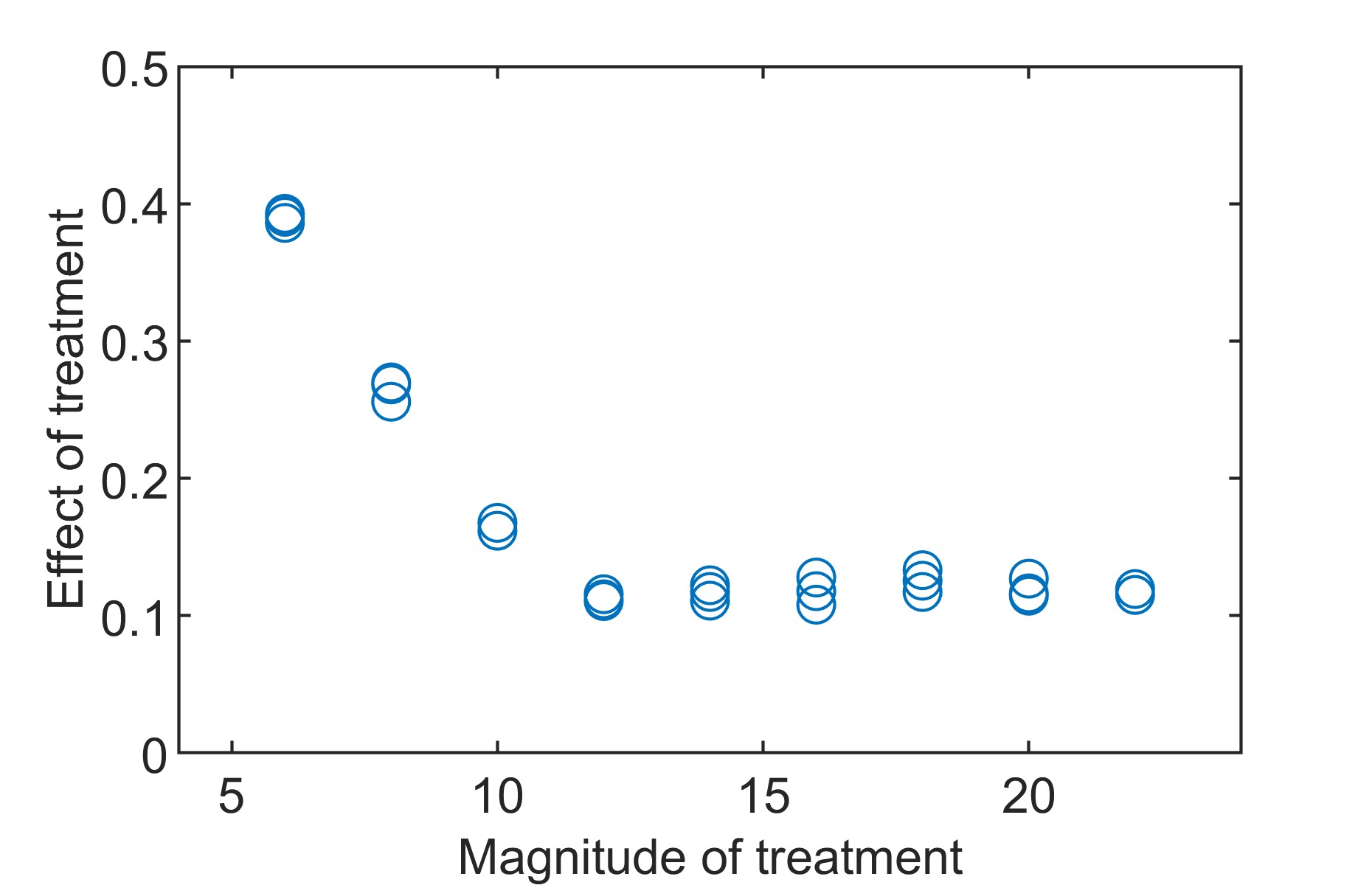}
    \caption{The reconstructed $\hat{\boldsymbol{\delta}}$ function.  }
    \label{fig7}
\end{figure}

\begin{figure}
    \centering
    \includegraphics[width=3.5 in]{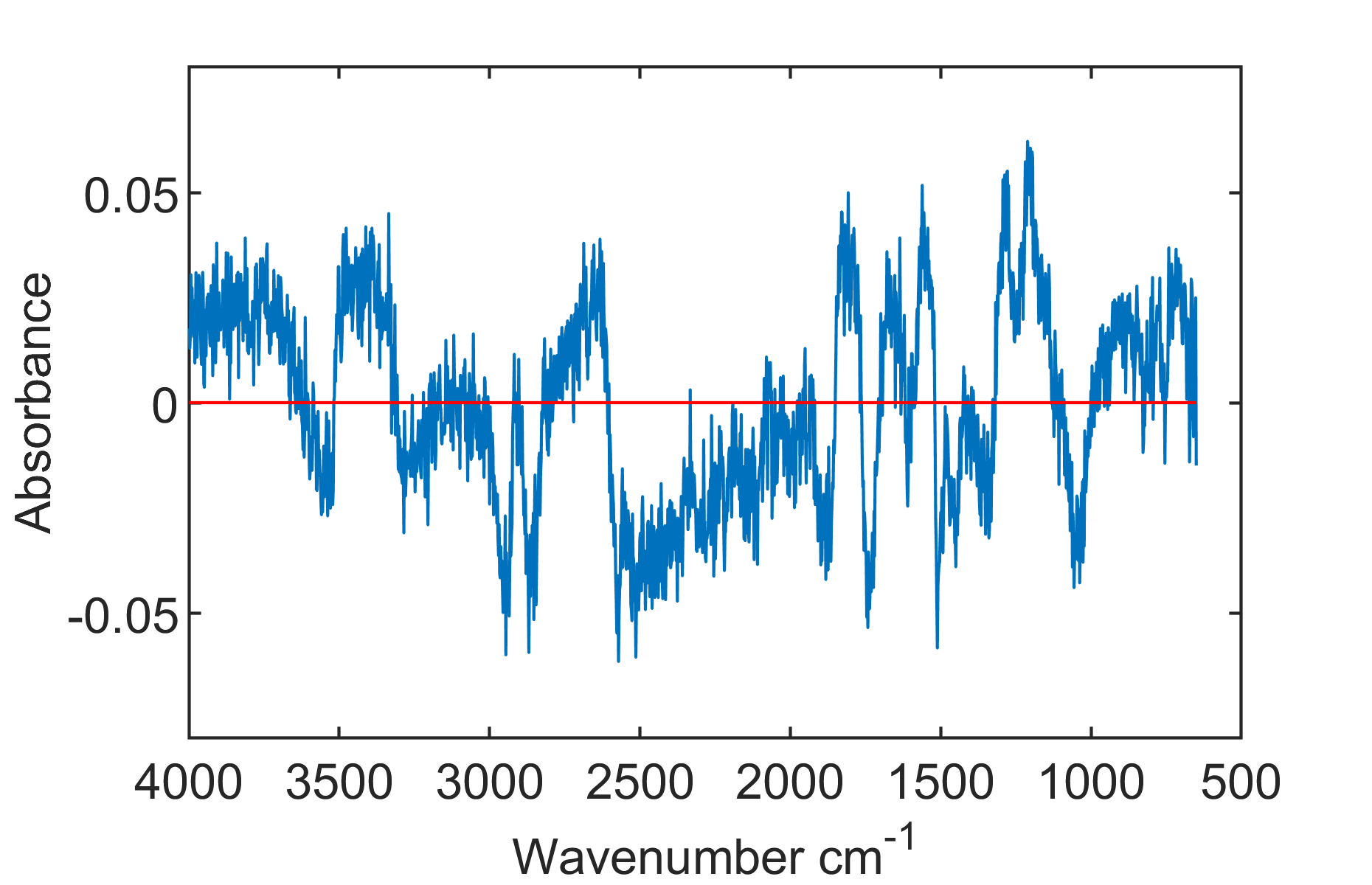}  
    \newline
    \caption{The signal $\tilde{\mathbf{g}}$  solved from \eqref{eq:step2constr}}
    \label{fig8}
\end{figure}

Then, we regard the pre-treatment template FTIR signal $\mathbf{x}_0$ as known and implement the proposed coordinate descent method on the simulated post-treatment signals to estimate the vector of effect $\boldsymbol{\delta}$ and the pattern of modification $\mathbf{g}$. After solving the problem (\ref{eq:step2constr}), we obtain the estimation of the vector of effect $\boldsymbol{\hat \delta}$ and component of the pattern of modification $\tilde{\mathbf{g}}$, as shown in Fig.~\ref{fig7} and Fig.~\ref{fig8}.  

In Fig.~\ref{fig7}, the estimated vector of effect $\widehat{\boldsymbol{\delta}}$ is plotted against the treatment magnitude. From this figure, it can be seen that the shape of  $\widehat{\boldsymbol{\delta}}$  is very similar to that of the true value $\boldsymbol{\delta}$ illustrated in Fig.~\ref{fig5}. It verifies that the accuracy of estimating the vector of effect.
Note that the vector of effect $\boldsymbol{\delta}$ is significantly different from the ground truth in Fig.~\ref{fig5}, which is primarily due to the constraints added when solving the optimization problem. It doesn't matter since what is of interest is the trend of the $\boldsymbol{\delta}$ rather than the magnitude. 
The variation in $\widehat{\boldsymbol{\delta}}$ is primarily caused by the estimation error of $\mathbf{a}_{1}, \mathbf{b}_{1}$, and the random error $\boldsymbol{\varepsilon}_1$. 


\begin{figure}
    \centering
    \includegraphics[width=3.5 in]{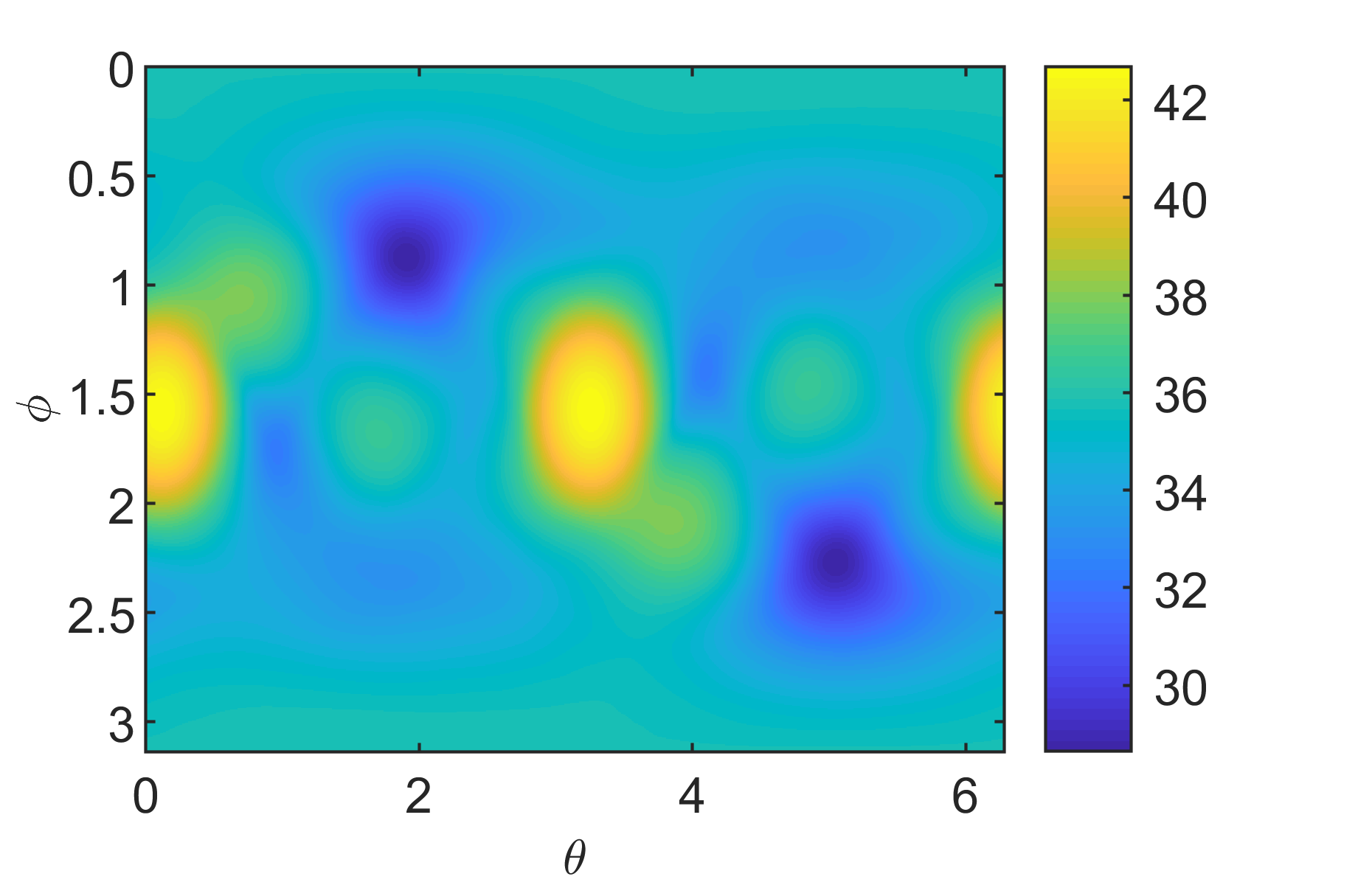}  
    \newline
    \caption{The heatmap of function $G(\theta, \phi),$ when $\theta \in[0,2 \pi], \phi \in[0, \pi]$  }
    \label{fig9}
\end{figure}

\begin{figure}
    \centering
    \includegraphics[width=3.5 in]{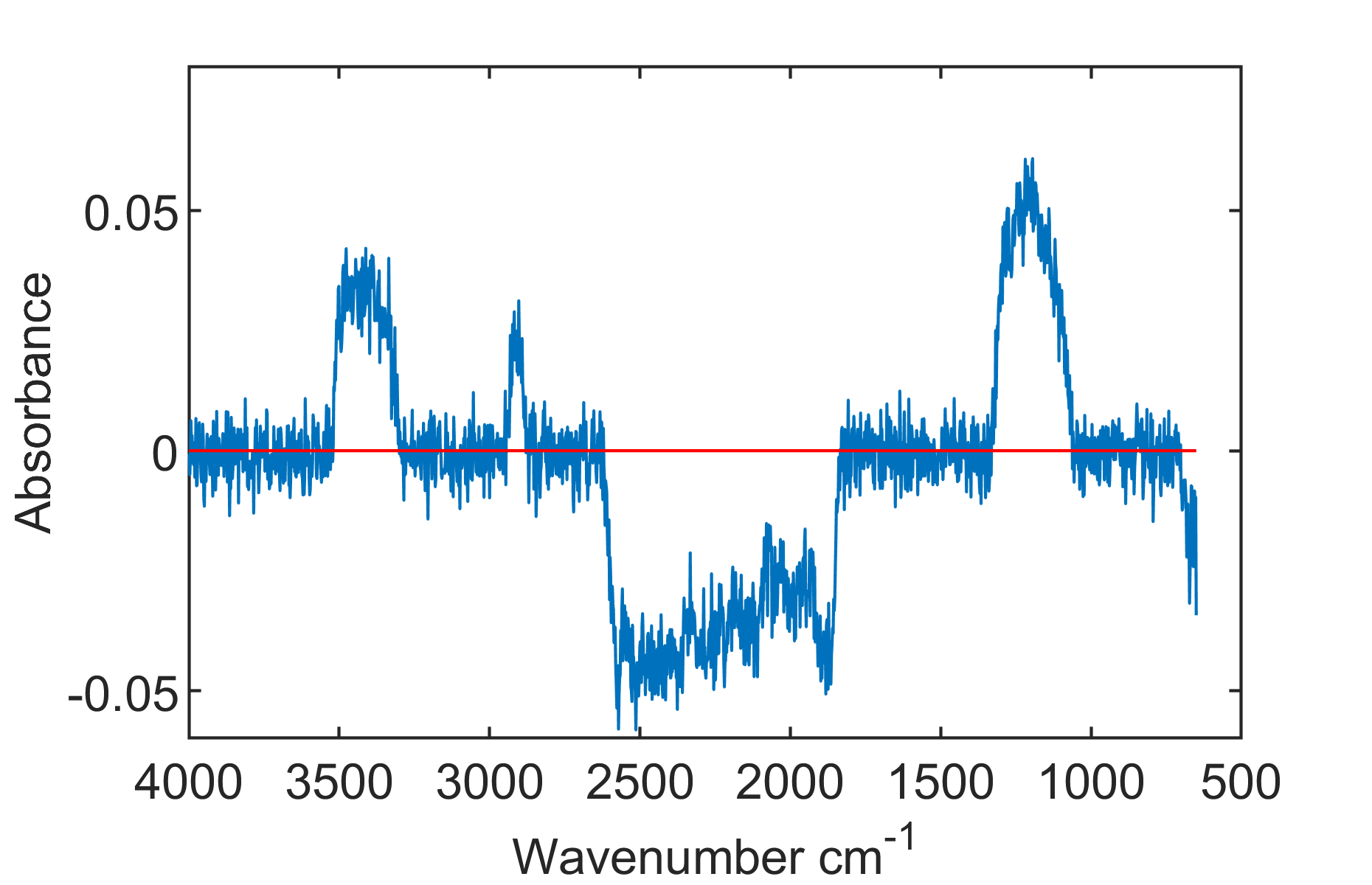}  
    \newline
    \caption{The reconstructed $\hat{\mathbf{g}}$  function. }
    \label{fig10}
\end{figure}

Fig.~\ref{fig8} illustrates the shape of the signal $\tilde{\mathbf{g}}$. To find the pattern of modification $\mathbf{g}$ with the best interpretability, we need to find $\theta^{*},~\phi^{*}$  that minimize the value of $G(\theta, \phi)$.  To understand the landscape of $G(\theta, \phi)$, we plot its values for all values $\theta\in[0,2\pi]$ and $\phi\in[0,\pi]$, as shown in Fig.~\ref{fig9}. From this figure, we can see that $G(\theta, \phi)$  is highly non-convex, and it has multiple local minima. Among these local minima, we select the one with  $\phi\approx 0$ or $\phi\approx \pi$, as we desire the function $\mathbf{g}$  be mainly determined by $\tilde{\mathbf{g}}$, and thus $\cos(\phi)$  be close to 1. We pick the local optima $\theta^{*}=1.8972$,~ $\phi^{*}=0.8741$, and the resulted vector of $\widehat{\boldsymbol{g}}$  is illustrated in Fig.~\ref{fig10}. From this result, the estimated $\widehat{\boldsymbol{g}}$  function in Fig.~\ref{fig10} is consistent with the ground truth of $\mathbf{g}$ in Fig.~\ref{fig6}. Our simulation study verifies the accuracy of the estimation of $\boldsymbol{g}$ as well. 
Note that the magnitude of the absorbance is significantly different from the ground truth in Fig.~\ref{fig6}. Similar to the reconstructed $\boldsymbol{\delta}$ function, such difference is believed to be caused by the constraints in solving the optimization problem. The shape of $\mathbf{\hat g}$ is close to $\mathbf{g}$ thereof.

\section{Case Study: Investigation of Real FTIR Spectra }
\begin{figure}
    \centering
    \includegraphics[width=3.5 in]{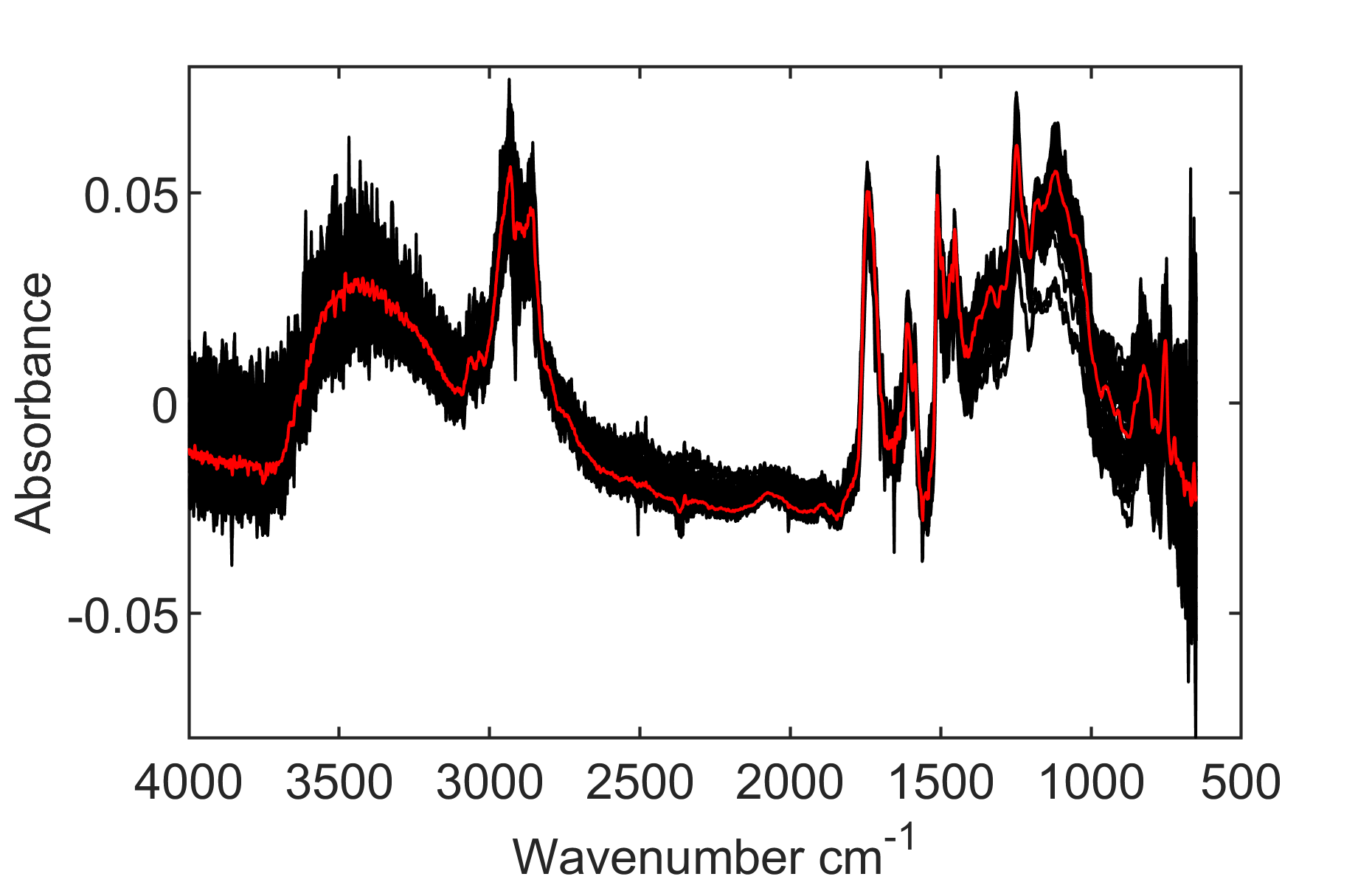} 
    \caption{ Corrected pre-treatment FTIR signals (black curves) and the template signal (red curve).}
    \label{fig11}
\end{figure}

\begin{figure}
    \centering
    \includegraphics[width=3.5 in]{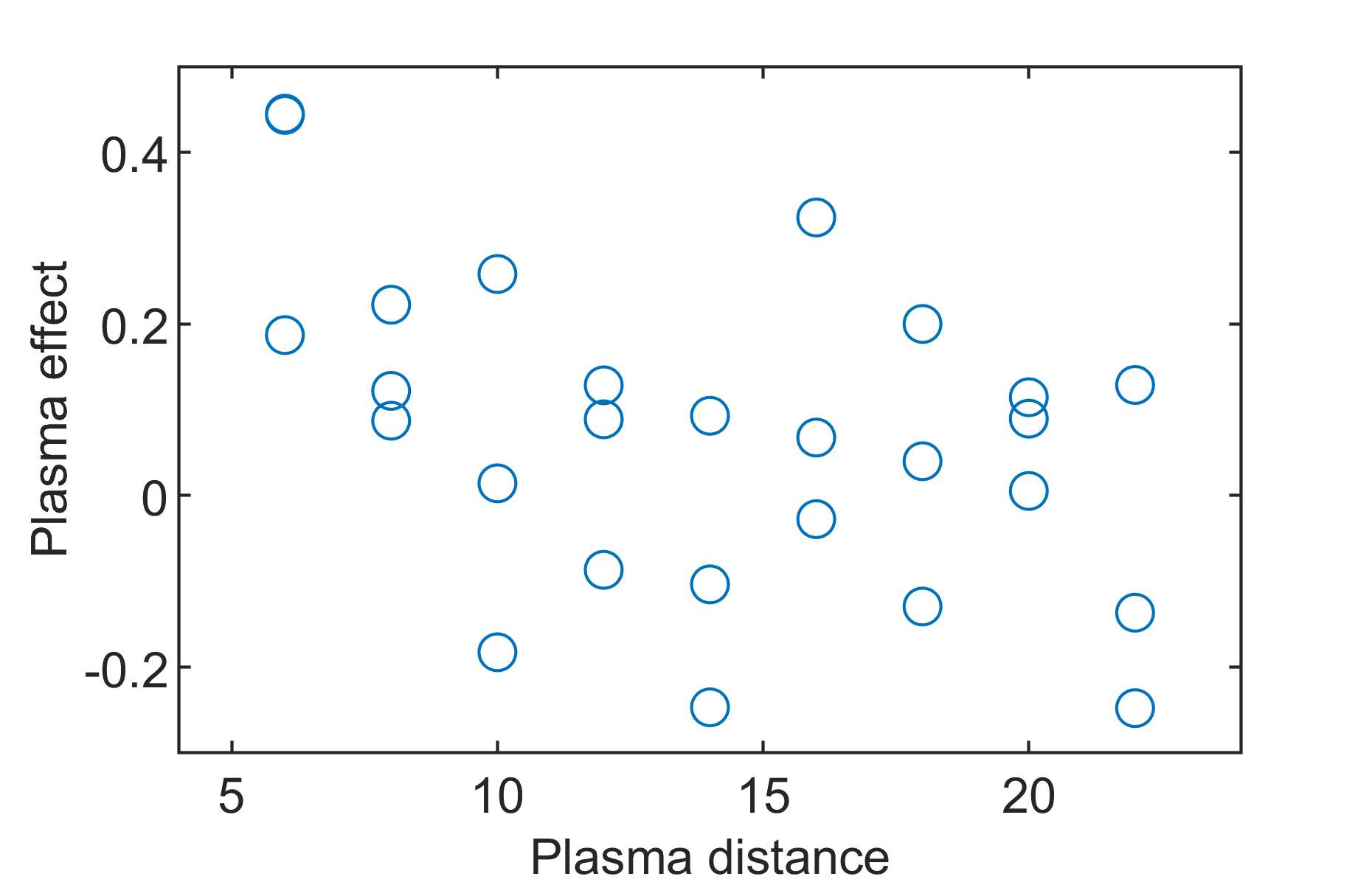}  
    \caption{ The plot of the effect of plasma exposure $\boldsymbol{\delta}$  versus the corresponding plasma height. }
    \label{fig12}
\end{figure}

The study is motivated by the engineering problem of understanding the effect of plasma exposure  on the CFRP panels when the plasma height varies. In this section, we further demonstrate how to use the method proposed in Section 2 to quantify the effect of plasma exposure, and discuss how the result benefit manufacturing engineering. We begin with introducing the setting of the experiment that collects the pre-exposure and post-exposure FTIR signals when the plasma height varies. Then we demonstrate our analysis procedure in detail, and discuss our findings. 

\subsection{Experimental Setup and Data Collection}
The CFRP sheet is the base material used in the experiment shown in Fig.~\ref{fig1}. In this experiment, we prepared 
eleven $ 1" \times 1"$ CFRP panels fabricated with HexPlyR M20 curing epoxy matrix from the same batch.

First, we measured the FTIR signal three times on each coupon before plasma treatment to obtain $n = 33$ pre-exposure FTIR spectra. In our experiment, the FTIR equipment used is 4300 Handheld FTIR from Agilent Company. On a small area on the sample, it captures the absorbance intensity of infrared light whose wavenumber (determined by frequency) is between 650 cm$^{-1}$ and 4000 cm$^{-1}$.
Within this range, each FTIR spectrum is represented by a signal containing $p=1798$ equidistant data points. 

To understand the effect of plasma treatment when the plasma height varies, these eleven sample coupons are processed with atmospheric press plasma with prescribed plasma heights $h = 22$mm, 20mm, ..., 2mm. 
Again we take three FTIR measurements on each coupon, with the same procedure as the pre-exposure measurement. 
We noted that the six post-exposure FTIR signals from the two coupons that underwent the plasma exposure with 2mm and 4mm plasma heights (denoted as black curves in Figure~\ref{fig2} (b)) are significantly different from the signals obtained from the other coupons, and investigated from the sample that the surfaces are carbonized due to the excessive heat caused by the plasma exposure. Therefore, these six FTIR measurements are eliminated, and  the post-exposure spectra contain 27 FTIR signals. 

\subsection{Implementation of the Proposed Method on Real Data}
Similar to the procedure of simulation on synthetic data, we implement the proposed methodology on real FTIR data collected from the experiment described in the last part. In Step 1, we estimate the template signal  $\mathbf{x}_{0}$ with the aligned pre-exposure FTIR signals by solving problem \eqref{eq:step1}, and the result is shown in Fig.~\ref{fig11}. The black curves are all aligned pre-exposure signals $\hat{c}_{0, i} \mathbf{x}_{0, i}+\hat{d}_{0, i} \mathbf{1}$, and the red curve is the template signal $\mathbf{x}_0$. Compared with the raw data seen in Fig.~\ref{fig2}, the pre-exposure signals are well aligned, and the template signal $\mathbf{x}_0$ is the representation of these curves.

\begin{figure}
    \centering
    \includegraphics[width=3.5 in]{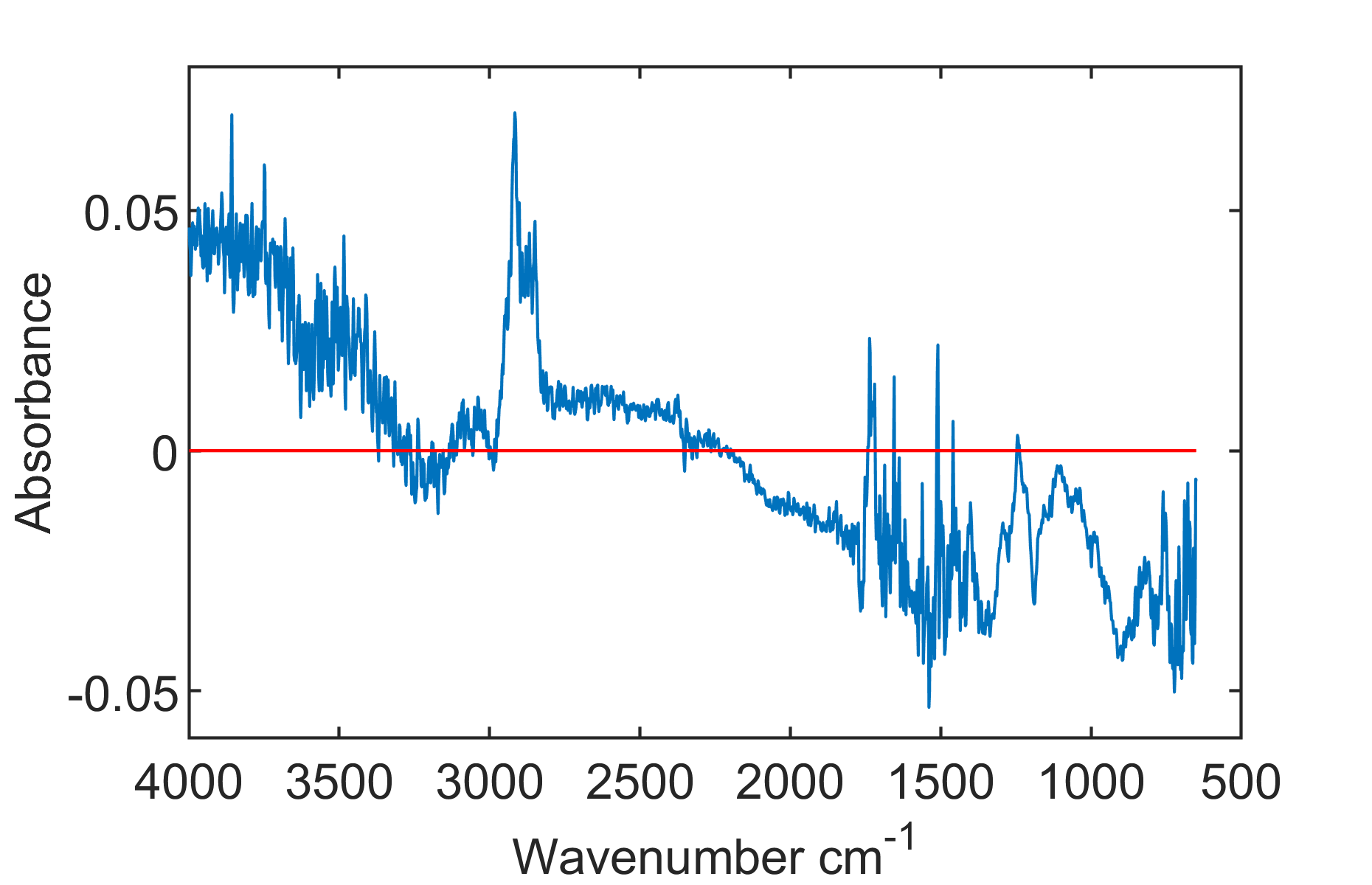}  
  \caption{ The signal  $\tilde{\mathbf{g}}$, as solved from the problem \eqref{eq:step2constr} }
    \label{fig13}
\end{figure}

Then, we implement the proposed coordinate descent method on the post-exposure signals to quantify the effect of plasma surface treatment. Similar to the process in the last section, we obtain the vector of effect  $\boldsymbol{\delta}$ and the signal $\tilde{\mathbf{g}}$ by solving the problem \eqref{eq:step2constr}, as shown in Fig.~\ref{fig12} and Fig.~\ref{fig13}. The vector of effect $\boldsymbol{\delta}$  is plotted against the plasma height in Fig.~\ref{fig12}. From this figure, we can see that the effect of the plasma treatment is positive when the plasma height is small, and gradually decreases as the plasma height increases, and generally becomes constant after the plasma height gets greater than 10mm. 
From the trend, we can see that a smaller plasma height tends to cause a more significant chemical change. This result confirms our previous experiment that the wettability of the CFRP material is higher when smaller plasma height is used, as long as the CFRP surface is not burnt (which is also mentioned in \cite{BTG2013CA}).

The result in Fig.~\ref{fig12} can provide valuable guideline to plasma parameter selection during CFRP surface preprocessing in aircraft maintenance. Although we know from the physical understanding that there is an upper bound of plasma height, under which the plasma treatment is effective, the shape of the vector $\boldsymbol{\delta}$ tells us that 10 mm is approximately the threshold, as larger plasma height will change the FTIR curve very little. In other words, the analysis gives us a relatively quantitative and persuasive understanding of the surface treatment effect on the FTIR measurements. 

\begin{figure}
    \centering
    \includegraphics[width=3.5 in]{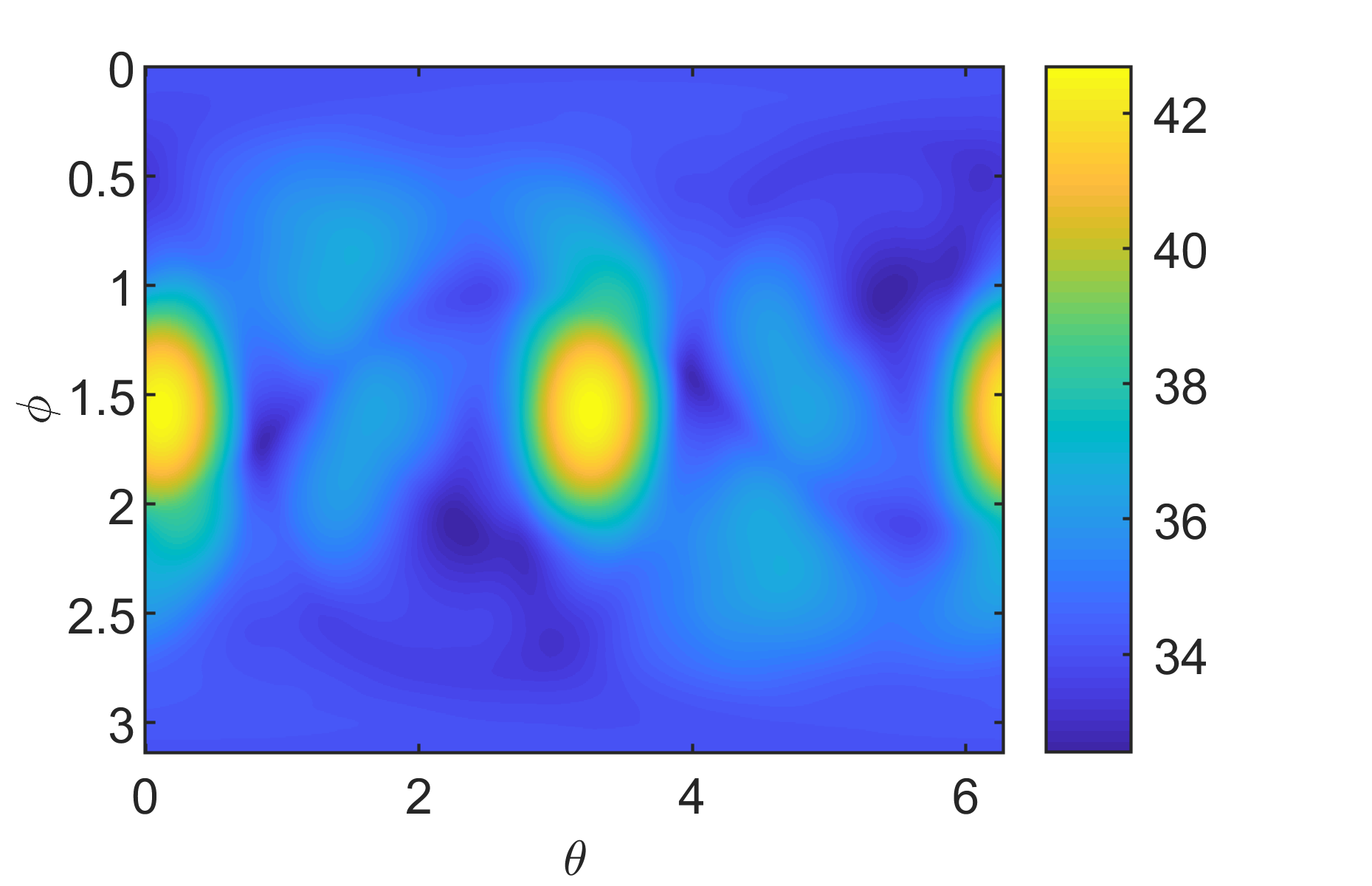}  
    \caption{The heatmap of function $G(\theta, \phi),$ when $\theta \in[0,2 \pi], \phi \in[0, \pi]$ }
    \label{fig14}
\end{figure}

\begin{figure}
    \centering
    \includegraphics[width=3.5 in]{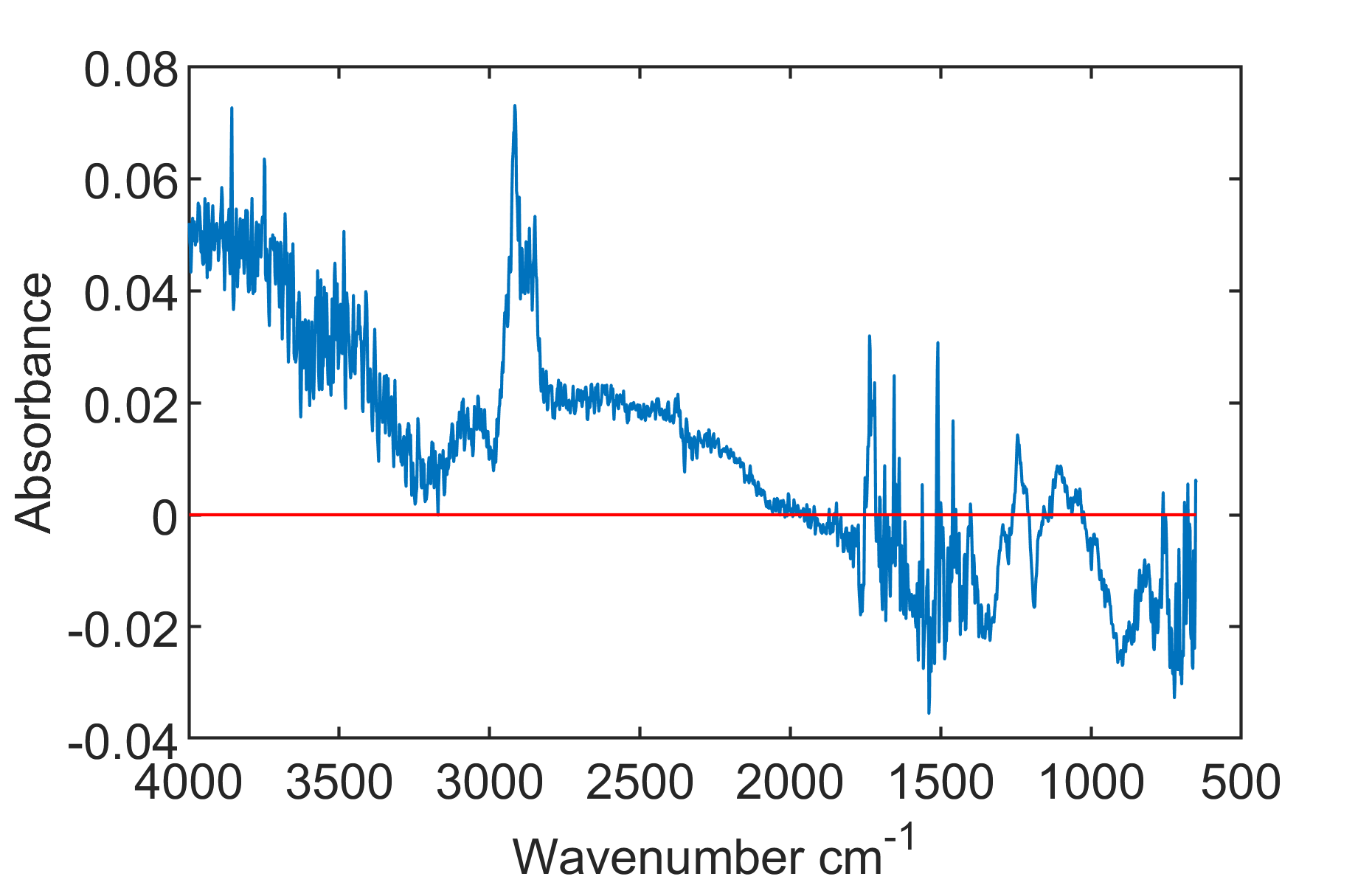}  
    \caption{The pattern of modification $\mathbf{g}$. }
    \label{fig15}
\end{figure}

Then we find the pattern of modification ${\mathbf{g}}$ that minimize the value of $G(\theta, \phi)$ based on $\tilde{\mathbf{g}}$ illustrated in Fig.~\ref{fig13}. The value of $G(\theta, \phi)$ is illustrated in Fig.~\ref{fig14}. We pick the local optima $\theta^{*}=0.0009,~ \phi^{*}=0.5053$, and the resulted vector of $\mathbf{g}$ is illustrated in Fig.~\ref{fig15}.

Recall that the pattern of modification $\mathbf{g}$  illustrates the effect of plasma exposure on all frequency bands. 
The change of FTIR gives  
comprehensive information on how the chemical bonds are changed as a result of the plasma exposure \cite{mukhopadhyay2002plasma,ctucureanu2016ftir,bassan2011light}. From Fig.~\ref{fig15}, we can see that in the wavenumber region of $2000-3000cm^{-1}$, $3200-4000cm^{-1}$, especially $2700-3000cm^{-1}$, the pattern of modification $\mathbf{g}$ has several significant positive values. It implies that the corresponding chemical bonds, including  O = C = O, N = C = O, N = N = N, N = C = N, N = C = S, C -- H, O -- H, and N -- H,  are created by plasma exposure, according to the IR absorption table \cite{IRsearch}. The pattern of modification $\mathbf{g}$ in the rest frequency regions tends to be around zero, which means that plasma does not change the chemical bonds whose characteristic wavenumbers are in those regions. Our result agrees with existing researches on the effects of plasma exposure on materials. For example, it is suggested in \cite{sharma2014carbon} that the plasma grafting to CF surface increased the proportion of oxygen atoms, and that the plasma modified carbon fiber shows a significant increase in oxygen and nitrogen concentration \cite{han2014evaluation}.

\section{Conclusion and Discussion}
In this article, we proposed a general framework to quantify the effects of certain treatment on the FTIR spectra subjecting to multiple uncertainties. With this framework, two types of uncertainties in the FTIR signals, offset shift and multiplicative error, were carefully addressed. In the two-step procedure, we first formulated a novel optimization problem to estimate the representative pre-treatment spectrum, and then formulated another optimization problem to obtain a pattern of modification $\mathbf{g}$ that reveals how the treatment affects the shape of the FTIR spectrum, as well as a vector $\boldsymbol{\delta}$ that describes the degree of change caused by different treatment magnitudes. 

This methodology is illustrated by a motivating example of quantifying the plasma treatment effect on the CFRP though FTIR measurements. In the case study, we understand the effective range of the plasma height from the vector of effects $\mathbf{\delta}$, and identified the affected chemical bounds from the pattern of modification $\mathbf{g}$. 
In future research, we can go one step further to map $\mathbf{g}$ to the change of constituents of the chemical compounds with the help of the FTIR spectra librarys \cite{IRsearch}. 
The knowledge on the modification of chemical components can shed some light on how the surface improvement technology change the chemical component of the material, which further indicate which chemicals shall be added or avoided to improve the surface quality before composite joining and repairing in aircraft manufacturing and maintenance.

Our technique is suitable and promising to analyze the data obtained from a wide range of spectral measurements, including ultraviolet-visible spectroscopy (UV-Vis), X-ray diffraction (XRD), Raman spectroscopy etc, when the background noise leads to uncertain offset and the uncertain signal level results in the multiplicative error. Typically, these uncertainties are of greater magnitude in real manufacturing environments than the lab conditions, due to the inexperienced operators and the uncontrolled surroundings. 
Therefore, the analytic framework proposed in this article also helps to broaden the scope of portable spectrometers, such as the hand-held FTIR devices. 

Also, the method in this article is applicable to quantify the effect a large class of surface treatment apart from the plasma treatment. 
The pattern of modification $\mathbf{g}$ reveals the general effect pattern of certain surface treatment method in a relatively large range of the treatment magnitude, which provides a better understanding of intrinsic reasons behind the treatment.
Recall that the surface treatment methods have been well-developed, including thermal treatment, wet chemical or electrochemical oxidation, plasma treatment, gas-phase oxidation, coating treatment, and so on. 
Applying and extending this method to solve more problems in the manufacturing industry is a direction for our further research. 


Finally, the methodology of this study can also be extended. For example, this study assumes that there is a single pattern of modification $\mathbf{g}$, which is the same under all levels of treatment effect. New methodologies can  be developed in future research, based on the assumptions that the pattern of modification is different for distinct levels of surface treatment. 


\bibliographystyle{ieeetr}
\bibliography{biolib.bib}

\ifdefined\draft

\else
\begin{IEEEbiography}[{\includegraphics[width=1in,height=1.25in,clip,keepaspectratio]{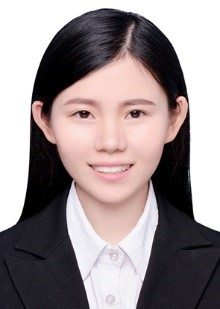}}] {Hongzhen Tian}
received the B.S. degree in Energy and Power Engineering System and Its Automation from the Xi’an Jiaotong University, Xi’an, China, in 2017. 

She is currently pursuing the Ph.D. degree with the H. Milton Stewart School of Industrial and Systems Engineering, Georgia Institute of Technology, Atlanta, GA, USA.

Miss Tian is a member of the Institute of Industrial and Systems Engineers (IISE). Her current research interests include statistical modeling and image analysis in engineering applications.

\end{IEEEbiography}

\begin{IEEEbiography}[{\includegraphics[width=1in,height=1.25in,clip,keepaspectratio]{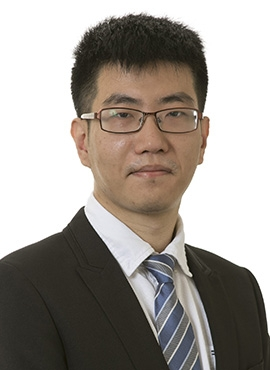}}] {Andi Wang}
received the B.S. degree from the School of Mathematical Sciences, Peking University, Beijing, China, in 2012. He received the Ph.D. degree from the Department of Industrial Engineering and Logistic Management, The Hong Kong University of Science and Technology, Hong Kong, in 2016. 

He is currently pursuing the Ph.D. degree with the H. Milton Stewart School of Industrial and Systems Engineering, Georgia Institute of Technology, Atlanta, GA, USA. His current research interests include data analytics and machine learning for the diagnostics, control and monitoring of complex engineering applications.
\end{IEEEbiography}

\begin{IEEEbiography}[{\includegraphics[width=1in,height=1.25in,clip,keepaspectratio]{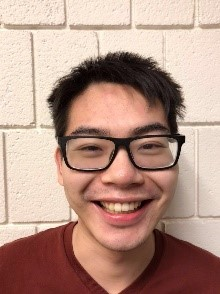}}] {Jialei Chen}
received the B.S. degree in theoretical and applied mechanics from the University of Science and Technology of China, Hefei, China, in 2016.

He is currently pursuing the Ph.D. degree with the H. Milton Stewart School of Industrial and Systems Engineering, Georgia Institute of Technology, Atlanta, GA, USA.

Mr. Chen is a member of the Institute of Industrial and Systems Engineers (IISE). His current research interests include physics-informed learning approaches, design and analysis of computer experiments, and medical image analysis.

\end{IEEEbiography}

\begin{IEEEbiography}[{\includegraphics[width=1in,height=1.25in,clip,keepaspectratio]{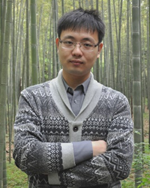}}] {Xuzhou Jiang}
received the B. S. degree in chemical engineering and  bioengineering from Zhejiang University, Hangzhou, China, in 2010, and the Ph. D. degree in mechanical engineering from Hong Kong University of Science and Technology, Hong Kong, China, in 2016.

He is currently a post-doctoral fellow in Georgia Tech Manufacturing Institute, Georgia Institute of Technology, Atlanta, GA, USA. His current research interests include additive manufacturing, printed electronics technologies, advanced sensing technologies and their applications in cell manufacturing, composite joining and repair, and smart materials.
\end{IEEEbiography}

\begin{IEEEbiography}[{\includegraphics[width=1in,height=1.25in,clip,keepaspectratio]{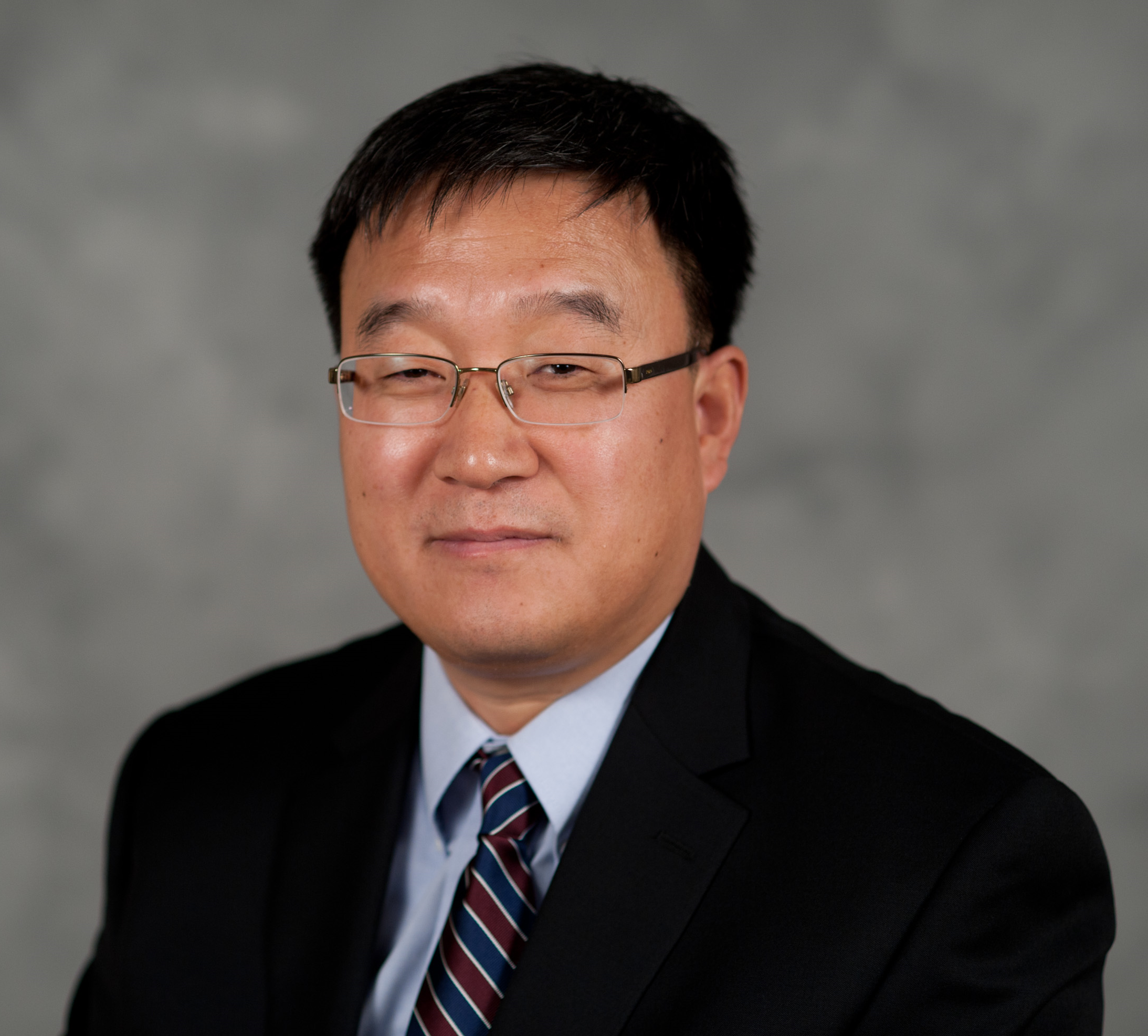}}] {Jianjun Shi}
received the B.S. and M.S. degrees in electrical engineering from the Beijing Institute of Technology, Beijing, China, in 1984 and 1987, respectively, and the Ph.D. degree in mechanical engineering from the University of Michigan, Ann Arbor, in 1992. 

Currently, he is the Carolyn J. Stewart Chair Professor in the H. Milton Stewart School of Industrial and Systems Engineering and George W. Woodruff School of Mechanical Engineering, Georgia Institute of Technology, Atlanta. His research interests include the fusion of advanced statistical and domain knowledge to develop methodologies for modeling, monitoring, diagnosis, and control for complex manufacturing systems.

Dr. Shi is a Fellow of the IIE, a Fellow of ASME, a Fellow of INFORMS, an academician of the International Academy for Quality, an elected member of the ISI, a life member of ASA, and a member of National Academy of Engineering (NAE).
\end{IEEEbiography}

\begin{IEEEbiography}[{\includegraphics[width=1in,height=1.25in,clip,keepaspectratio]{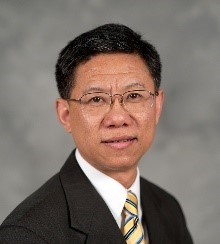}}] {Chuck Zhang}
received the B.S. and M.S. degrees in mechanical engineering from Nanjing University of Aeronautics and Astronautics, Nanjing, China, in 1984 and 1987, respectively, an MS degree in Industrial Engineering from the State University of New York at Buffalo in 1990, and the Ph.D. degree in industrial engineering from the University of Iowa, Iowa City, IA, USA, in 1993.

Dr. Zhang is currently a Harold E. Smalley Professor with the H. Milton Stewart School of Industrial and Systems Engineering at the Georgia Institute of Technology, Atlanta, GA, USA. He has authored over 190 refereed journal articles and 210 conference papers. He also holds 24 U.S. patents. Dr. Zhang is a fellow of Institute of Industrial and Systems Engineers (IISE). His current research interests include additive manufacturing (3-D printing and printed electronics), advanced composite and nanocomposite materials manufacturing, and bio-manufacturing.
\end{IEEEbiography}

\begin{IEEEbiography}[{\includegraphics[width=1in,height=1.25in,clip,keepaspectratio]{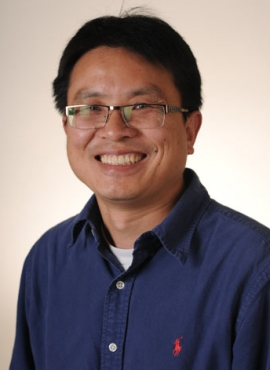}}] {Yajun Mei} received a B.S. in Mathematics from Peking University in P.R. China, and a Ph.D. in Mathematics with a minor in Electrical Engineering from the California Institute of Technology. He has also worked as a postdoc in Biostatistics for two years in the Fred Hutchinson Cancer Research Center in Seattle, WA. 

Dr. Mei's research interests include change-point problems and sequential analysis in Mathematical Statistics; sensor networks and information theory in Engineering; as well as longitudinal data analysis, random effects models, and clinical trials in Biostatistics.
\end{IEEEbiography}

\begin{IEEEbiography}[{\includegraphics[width=1in,height=1.25in,clip,keepaspectratio]{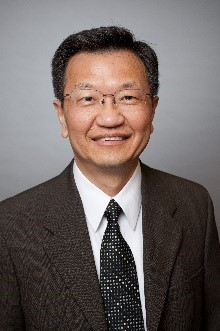}}] {Ben Wang}
is Executive Director of the Georgia Tech Manufacturing Institute, Atlanta. He is Gwaltney Chair in Manufacturing Systems in the School of Industrial and Systems Engineering, Georgia Institute of Technology, where he is also Professor in the School of Materials Science and Engineering. He has authored more than 260 refereed journal papers journals and is a Co-Inventor of 35 patents or patent applications. With a primary research interest in applying emerging technologies to improve manufacturing competitiveness, he specializes in process development for affordable composite materials and is widely acknowledged as a pioneer in the growing field of nanomaterials. His current research interests include high performance and affordable composites, which is already changing product innovations worldwide.

Dr. Wang is a fellow of the Institute of Industrial Engineers, Society of Manufacturing Engineers, and Society for the Advancement of Materials and Process Engineering. He chairs the National Materials and Manufacturing Board of the National Academies of Sciences, Engineering and Medicine.
\end{IEEEbiography}
\fi



\end{document}